\documentclass[aps,prd,nofootinbib,floatfix,preprintnumbers,tightenlines,11pt,superscriptaddress,abstract=on]{revtex4-1}

\usepackage{amsfonts}
\usepackage{amsmath}
\usepackage{amssymb}
\usepackage{bm}
\usepackage{graphicx}
\usepackage{cancel}
\usepackage[pdftex]{color}
\usepackage[sort&compress]{natbib}
\usepackage[colorlinks=true,linkcolor=blue,filecolor=blue,urlcolor=blue,citecolor=blue,pdftex,plainpages=false]{hyperref}

\mathchardef\mhyphen="2D

\begin{document}

\title{The Role of Lattice QCD in Searches for Violations of Fundamental Symmetries and Signals for New Physics}
\collaboration{USQCD Collaboration}
\noaffiliation
\author{Vincenzo Cirigliano (Editor)}\email{cirigliano@lanl.gov }
\affiliation{
	Theoretical Division T-2, Los Alamos National Laboratory, Los Alamos, NM 87545, 		USA}
\author{Zohreh Davoudi (Editor)}\email{davoudi@umd.edu}
\affiliation{
 	Department of Physics and Maryland Center for Fundamental Physics, 
University of Maryland, College Park, MD 20742, USA}
\affiliation{
 	RIKEN Center for Accelerator-based Sciences, Wako 351-0198, Japan}

\author{Tanmoy Bhattacharya}
\affiliation{
	Theoretical Division T-2, Los Alamos National Laboratory, Los Alamos, NM 87545, 		USA}
\author{Taku Izubuchi}
\affiliation{
 	Physics Department, Brookhaven National Laboratory, Upton, NY 11973, USA}
\affiliation{
 	RIKEN-BNL Research Center, Brookhaven National Lab, Upton, NY, 11973, USA}
\author{Phiala E. Shanahan}
\affiliation{
 	Center for Theoretical Physics, 
 	Massachusetts Institute of Technology, 
 	Cambridge, MA 02139, USA}
\author{Sergey Syritsyn}
\affiliation{
	Department of Physics and Astronomy, Stony Brook University,
	Stony Brook, NY 11794}
\affiliation{
 	RIKEN-BNL Research Center, Brookhaven National Lab, Upton, NY, 11973, USA}
\author{Michael L. Wagman} 
\affiliation{
 	Center for Theoretical Physics, 
 	Massachusetts Institute of Technology, 
 	Cambridge, MA 02139, USA}


\begin{abstract}

\
\

This document is one of a series of whitepapers from the USQCD collaboration. Here, we discuss opportunities for Lattice Quantum Chromodynamics (LQCD) in the research frontier in fundamental symmetries and signals for new physics. LQCD, in synergy with effective field theories and nuclear many-body studies, provides theoretical support to ongoing and planned experimental programs in searches for electric dipole moments of the nucleon, nuclei and atoms, decay of the proton, $n$-$\overline{n}$ oscillations, neutrinoless double-$\beta$ decay of a nucleus, conversion of muon to electron, precision measurements of weak decays of the nucleon and of nuclei, precision isotope-shift spectroscopy, as well as direct dark matter detection experiments using nuclear targets. This whitepaper details the objectives of the LQCD program in the area of Fundamental Symmetries within the USQCD collaboration, identifies priorities that can be addressed within the next five years, and elaborates on the areas that will likely demand a high degree of innovation in both numerical and analytical frontiers of the LQCD research. 

\end{abstract}

\maketitle

\nopagebreak[4]

  \tableofcontents
\section*{Executive summary}
 \label{sec:exec}
\noindent
In 2018, the USQCD collaboration's Executive Committee organized several subcommittees to recognize future opportunities and formulate possible goals for lattice field theory calculations in several physics areas.  The conclusions of these studies, along with community input, are presented in seven whitepapers~\cite{Bazavov:2018qcd,Brower:2018qcd,Kronfeld:2018qcd,Detmold:2018qcd,Joo:2018qcd,Lehner:2018qcd}. The current whitepaper concerns the role of LQCD in the research frontier in fundamental symmetries and signals for new physics.

Precision tests of the Standard Model (SM) of particle physics, if they reveal any discrepancy with the model's predictions, provide a window into the physics beyond the SM. This would have a profound impact on our understanding of how nature works at the smallest distances and highest energies -- scales far beyond the reach of particle colliders of today and of the future. The U.S. Nuclear Science Advisory Committee (NSAC) promoted research in Fundamental Symmetries and Neutrino Physics to one of the primary thrusts of the field of Nuclear Physics (NP) in the 2007 Long Range Plan~\cite{Aprahamian:2007}. NSAC further enhanced and extended its support for such a program in the 2015 Long Range Plan~\cite{Geesaman:2015fha}, including making the strongest recommendation for investment in a tonne-scale experiment that will search for lepton-number violation in the neutrinoless double-$\beta$ ($0\nu\beta\beta$) decay of a nucleus. Other areas in which the U.S. NP and High Energy Physics (HEP) programs have invested substantially, and in several cases is leading the international effort, include searches for CP violation in electric dipole moments (EDMs) of the nucleon, nuclei and atoms, baryon-number nonconservation through proton decay, baryon-number minus lepton-number nonconservation through $n$-$\overline{n}$ oscillations, lepton-number nonconservation through $0\nu\beta\beta$ decay of a nucleus, and lepton-flavor nonconservation through muon to electron conversion. Further, precision tests of the SM through weak decays of neutrons and nuclei and isotope-shift spectroscopy are ongoing efforts. 
Furthermore, a devoted experimental effort towards direct detection of dark matter constrains many new physics scenarios, and accurate theoretical calculations of the cross sections of various dark-matter candidates with experimental nuclear targets are needed to reliably interpret experimental results.

All of these exciting frontiers share a common theme: new heavy particles and interactions could violate (approximate) symmetries of the SM and lead to new physics in low-energy observables accessible to high-intensity experiments involving nucleons and nuclei, even in scenarios where the energy scale associated with the symmetry violation is inaccessible in collider experiments. Constraining new-physics models requires matching, e.g., the observed rate of the experimental process, to new high-scale interactions whose effects are encoded in a variety of higher-dimensional operators present in the low-energy SM theory. High-scale interactions are expressed at the quark-(and lepton-)level, and nonperturbative calculations of the strong dynamics of quarks and gluons in the quantum chromodynamics (QCD) sector of the SM are needed in order to match theoretical predictions to low-energy observables involving nucleons and nuclei. LQCD, as the most reliable method to provide such an input, has long established its role in supporting the HEP experimental program, and is now pushing the boundaries of challenging calculations relevant for both NP and HEP in the Fundamental Symmetries program.

This whitepaper describes several areas in which LQCD has been, or will be, playing a major role, enumerates the challenges ahead, and identifies specific calculations that can have the highest impact on the larger program and which would complement nuclear and high-energy research in the U.S. and worldwide. In particular, the planned studies have been categorized as: i) \emph{straightforward}: those which will reach the required level of precision with the continuation of current hardware, software and personnel support in the next 5 years, ii) \emph{challenging}: those which have just started now, but the completion of which will require at least an order of magnitude increase in computational resources that will be available in the coming Exascale era, and iii) \emph{extremely challenging}: those that demand new ideas and paradigms in theory, algorithm and computation, but are still in reach of the LQCD community, granted its continual creativity and persistence.

  \section{Introduction}
  \label{sec:intro}
\noindent
Overwhelming observational evidence in nature points to the existence of new degrees of freedom that go beyond the SM of particle physics. Beyond-the-Standard-Model (BSM) scenarios aim to explain the matter dominance over antimatter in Universe, the nature of massive but extremely light neutrinos, the extreme hierarchy of scales associated with electroweak and gravity forces, and the nature of dark matter. The Large Hadron Collider (LHC) at CERN has made a major discovery by observing the last predicted particle within the SM, the Higgs boson~\cite{Aad:2015zhl}, but no evidence for BSM particles or interactions has been discovered as yet from LHC results, including runs with a total beam energy of 13 TeV. Motivated by a complementary discovery potential, the case is now stronger than ever for ``Intensity Frontier'' investigations. In the intensity-frontier experiments, the goal is shifted from attempting to produce ``on-shell'' BSM particles in high-energy collision, to detecting their effects when they are produced as ``off-shell'' intermediate states in low-energy experiments. While the sensitivity of collider-based investigations at the ``Energy Frontier'' is limited to few to tens of TeVs with current technology, Intensity Frontier investigations are sensitive to several types of BSM physics that could occur at scales orders of magnitude higher than those reached in colliders. With such a significant discovery potential, a vigorous experimental program in the U.S. and around the world has formed over the past few decades to search for the violations of fundamental symmetries of the SM, and for the deviation of given observables from their calculated SM values. Not only will a non-null result be a smoking gun for new physics, but also it can be used to test specific BSM models of high-scale physics.

BSM scenarios may involve new particles and interactions that originate at very high-energy scales, or alternatively may involve new physics at relatively low scales with extremely feeble couplings~\cite{Alexander:2016aln}. This whitepaper focuses on the former class of BSM scenarios, for which an effective field theory (EFT) framework can be adopted to encompass all SM extensions that respect the SM gauge group and have the same low-energy particle content as the SM~\cite{Weinberg:1979sa,Buchmuller:1985jz,Grzadkowski:2010es}. One challenge that often needs to be dealt with in carrying out the program in Fundamental Symmetries is that the SM processes that are sensitive to new physics often involve hadrons and nuclei as initial, final, and/or intermediate states at low energies, demanding that SM and \emph{ab initio} quantum many-body calculations be performed in a highly nonperturbative regime.

\begin{figure}
\centering
\includegraphics[width=0.9\columnwidth]{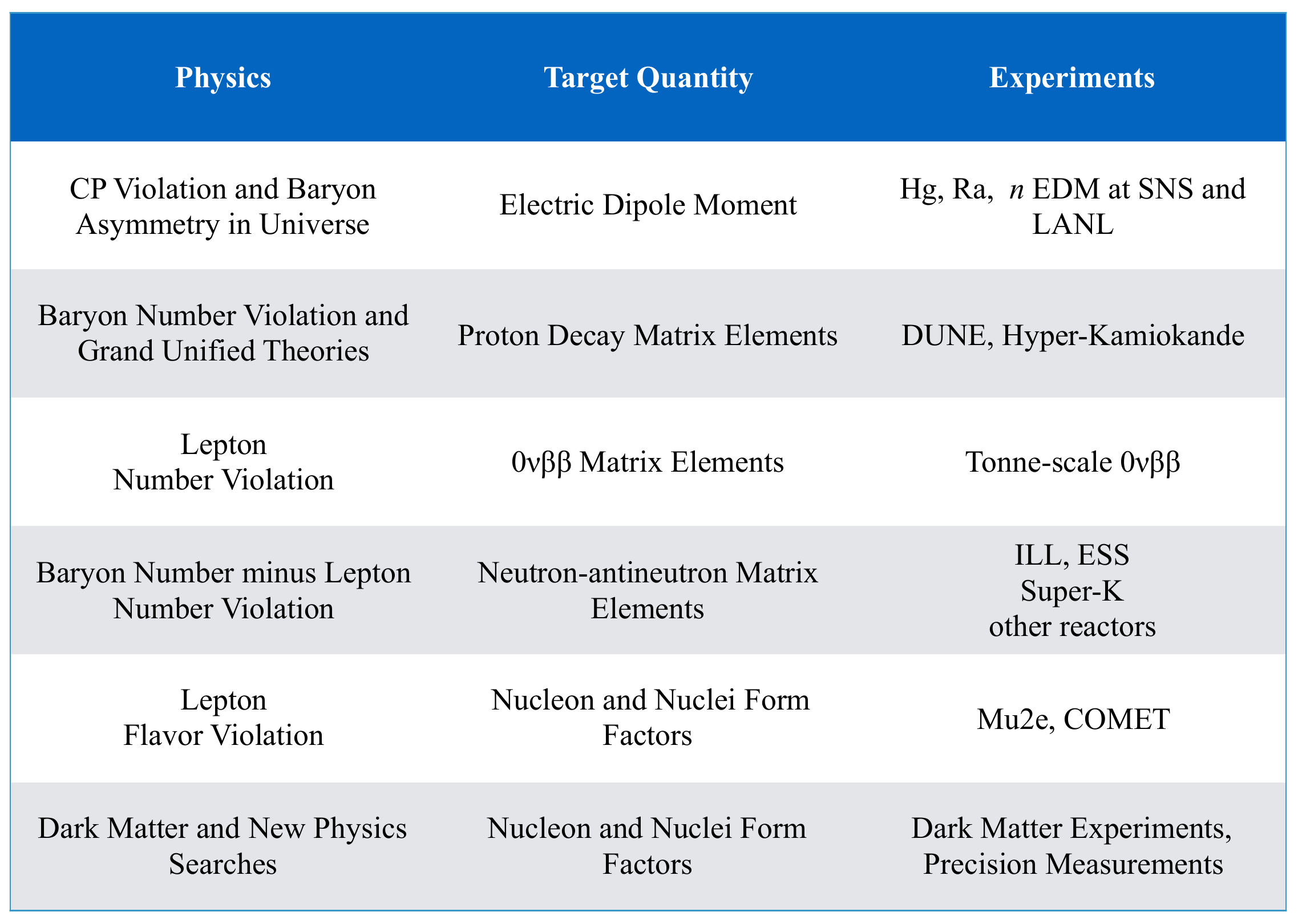}
\caption{Summary of primary Fundamental Symmetries subjects discussed in this whitepaper.}
\label{fig:resources}
\end{figure}

The Department of Energy (DOE)-funded USQCD collaboration, as the unified collaboration of the majority of the LQCD collaborations in the U.S., has long identified the role of LQCD studies in enabling and advancing research at the Intensity Frontier. The significant results produced by the collaboration in the past two decades in constraining the Cabibbo-Kobayashi-Maskawa (CKM) matrix quark-mixing matrix of the SM, and in constraining the quantities relevant to tests of CP violation in the Kaon sector, mark major accomplishments for the LQCD community, and signify the essential role of LQCD in complimenting both experiment and theory effort in HEP research. Such a role has been identified in other sectors in which there is, or is expected to be, hints of deviation from the SM and of new physics. In particular, compared to the year 2013 when the Collaboration's previous whitepapers were released, in 2018 the Collaboration has dedicated three separate whitepapers to the role of LQCD in research at the Intensity Frontier:\footnote{Current and future opportunities in direct studies of the BSM models that complement collider searches of new physics, and require the nonperturbative tool of lattice field theory to simulate potential strong dynamics, are detailed in another companion whitepaper by the Collaboration, see~\cite{Brower:2018qcd}.}
\begin{itemize}
\item[]{
\emph{i) Opportunities for LQCD in quark and lepton flavor physics:} Given the persisting tensions between theoretical predictions based on the SM and and experimental measurements in B meson decays at the LHC-B and the anomalous magnetic moment of the muon at the Brookhaven National Laboratory, and the continuing improvement in determinations of the CP violating parameters in Kaon and recently the heavy-meson sectors, there is a pressing need for reliable SM predictions of corresponding quantities from LQCD, to a precision level that is comparable to experimental determinations. Progress made by the USQCD collaboration in all these areas, the challenges ahead, and the plans forward are detailed in a companion whitepaper, see \cite{Lehner:2018qcd}.}
\item[]{\emph{ii) LQCD and neutrino-nucleus scattering:} The large experimental enterprise in neutrino physics, in particular the upcoming Deep Underground Neutrino Experiment (DUNE) at Fermilab aims to shed light on CP violation in the lepton sector. However, reliable determinations of the Pontecorvo-Maki-Nakagawa-Sakata (PMNS) lepton-mixing matrix require accurate knowledge of neutrino-nucleus interactions. The neutrino-nucleon and neutrino-nucleus scattering cross sections are hence of crucial value, as otherwise the neutrino energy deposit in the nuclear detector can not be reliably calculated. LQCD, in close collaboration with the nuclear structure community, will play a fundamental role in constraining these cross sections, an opportunity that has been detailed in a companion whitepaper, see \cite{Kronfeld:2018qcd}.}
\item[]{\emph{\textbf{iii) The role of LQCD in searches for violations of fundamental symmetries and signals for new physics:}} The violations of the symmetries protected in the SM go beyond the CP and lepton universality violation explored in lepton and flavor factories. A number of areas with significant growth in interest and potential include searches for CP violation in EDMs of the nucleon, nuclei and atoms, baryon-number nonconservation through proton decay, baryon-number minus lepton-number nonconservation through $n$-$\overline{n}$ oscillations, lepton-number nonconservation through the $0\nu\beta\beta$ decay of a nucleus, and lepton-flavor nonconservation through muon to electron conversion. Further, precision tests of the SM through weak decays of the nucleon and of nuclei and through isotope-shift spectroscopy are ongoing efforts. Furthermore, a devoted experimental effort in the direct detection of dark matter has been constraining new physics scenarios, but requires reliable theoretical calculations of the cross section of various dark matter candidates with the nuclear targets used in experiments. Given the strong dynamics involved in theoretical predictions based on the SM, LQCD again is playing an important role. This whitepaper details this program within the USQCD collaboration, and elaborates on the areas that will likely demand a high degree of innovation in both numerical and analytical frontiers of LQCD research. The table in Fig.~\ref{fig:resources} summarizes the quantities of interest for the purpose of this whitepaper, along with the physics objectives and the experimental programs they correspond to.}
\end{itemize}
Research in Fundamental Symmetries has became one of the primary research thrusts in NP in the U.S. following a 2007 Nuclear Science Advisory Committee (NSAC) Long Range Plan. A community whitepaper in 2015~\cite{Aprahamian:2015qag} explains that \emph{``...this development reflected a recognition that the subfield is an integral part of the NP scientific mission, that it provides unique opportunities for obtaining results having far-reaching significance, and that this potential is highly complementary to that of related disciplines, such as HEP and cosmology.''} A more recent NSAC Long Range Plan in 2015 once again emphasizes the continuation of support for research in Fundamental Symmetries, as highlighted by their highest recommendation for a major new investment in the U.S. NP experimental program, namely a tonne-scale experiment to search for a $0\nu\beta\beta$ decay, with the acknowledgment of the essential role of theory in the optimal design and in interpretation of the future result of such an experiment.
\begin{figure}[!t]
\centering
\includegraphics[width=0.95\columnwidth]{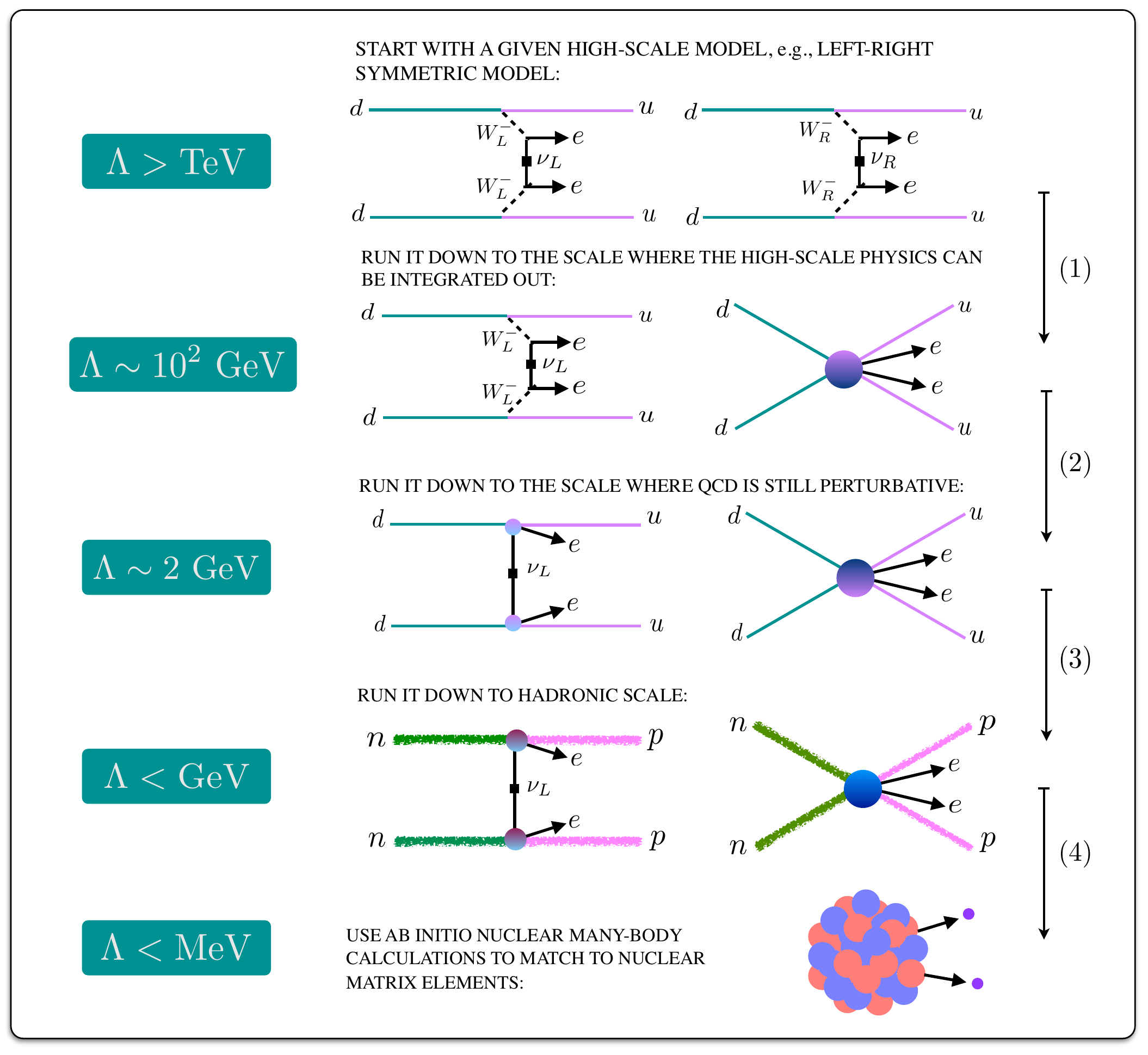}
\caption{The role of LQCD in searches for new physics is demonstrated through the example of $0\nu\beta\beta$ decay. To plan the design of new experiments, and to interpret any potential observation, theorists will need to apply a multi-scale matching program to connect the observed rate in heavy isotopes to the expectation from lepton-number violating models at a high scale, see e.g., Ref.~\cite{Cirigliano:2018yza}. LQCD is the only reliable method that enables taking step (3) in the chart, bridging between perturbative and nonperturbative QCD regimes, and providing new LECs of nuclear EFTs in the few-body sector that can be fed into \emph{ab initio} nuclear structure calculations of the isotopes of relevance to experiment.
}
\label{fig:0vbbchart}
\end{figure}

To understand the role of LQCD in the success of this program, one must recognize the complexity of connecting observations to the new-physics models one aims to constrain and the wide range of scales involved. The example of $0\nu\beta\beta$ decay demonstrates this complexity to the full degree (see also Sec.~\ref{sec:0vbb} for further detail on the science motivation and prospects for this program), and is a representative of the procedure involved in almost all frontiers in Fundamental Symmetries research~\cite{Cirigliano:2013lpa}. As illustrated in Fig.~\ref{fig:0vbbchart}, a multitude of lepton-number violating models at a hight scale (at and beyond TeV scale) can be proposed to explain a potential observation of this decay in a heavy isotope, such as Germanium $^{76}$Ge, where the typical nuclear scale is a few MeVs. In order to interpret the results within a given model and possibly differentiate between various models, it is necessary to perform a thorough matching between physics at the high scale and that at a low scale. Consider the left-right symmetric model, which extends the SM to contain a heavy right-handed neutrino and a corresponding heavy boson, $W_R$. In such a scenario, a $0\nu\beta\beta$ decay can occur not only as a result of the exchange of a light left-handed Majorana neutrino that converts two down quarks of the SM to two up quarks (the left column of the figure), but also a right-handed Majorana neutrino inducing the same transition (the right column of the figure). One then considers a low-energy EFT in which all of these ``heavy'' new mediators have been integrated out and relates the low-energy effective couplings to the high-energy new-physics parameters through the renormalization group (RG). Running down to below the electroweak symmetry breaking scale, the SM W bosons can be integrated out as well, leaving effective six-fermion interactions involving SM quarks and leptons. When running the scale down to the chiral symmetry breaking scale, the effective degrees of freedom are no longer quarks and gluons but rather hadrons and nucleons. To connect the effective couplings, or in turn the ``matrix elements'' (MEs) of the operators between nucleonic states to that of the higher scale requires a LQCD computation. Such calculations constrain the low-energy constants (LECs) of proper EFTs in the few-nucleon sector, which can then be input into \emph{ab initio} nuclear many-body calculations of the heavy isotopes used in experiments. It is important to note that the nature of such a nonperturbative calculation depends upon the high-scale scenario considered. In the scenario in which a light Majorana neutrino induces the decay, the neutrino can not be integrated out even at the nuclear energy scale, hence demanding complex calculations of a long-range nonlocal ME both at the QCD step and at the nuclear EFT and nuclear structure steps. As mentioned above, such a complex matching program is common to the majority of the research areas discussed in this whitepaper, up to technical differences, and is a prominent example of the need for a strong synergy between different subcommunities within HEP and NP.

  \section{CP violation and electric dipole moment of nucleon and nuclei}
  \label{sec:edm}
\begin{figure}
  \includegraphics[scale=0.675]{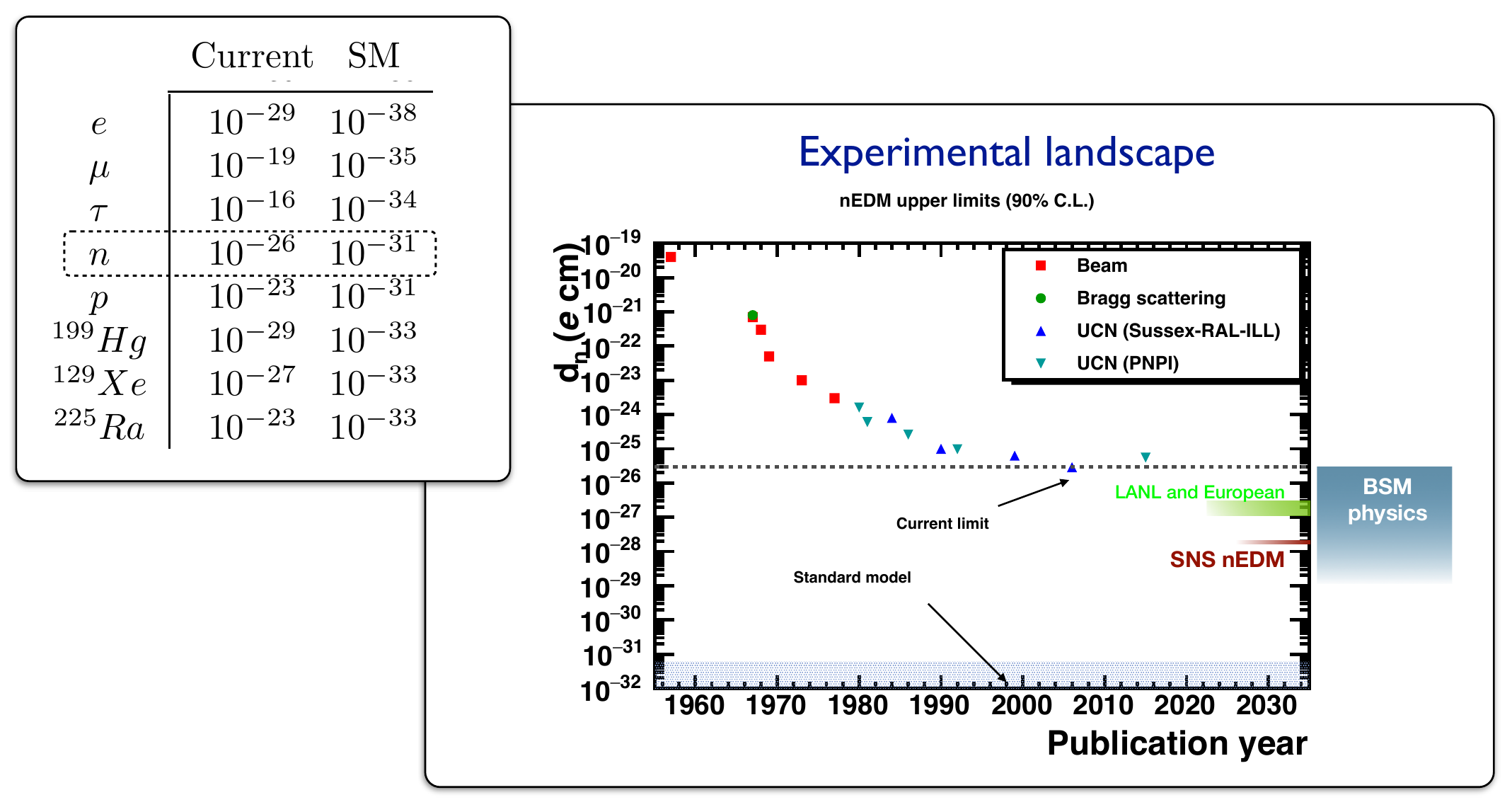}
  \caption{The table shows approximate limits (in units of $e\cdot$cm) on EDMs of various systems and their predictions from the  CP-violation within the SM through the CKM mechanism~\cite{Chupp:2017rkp}. The plot indicates that progress has slowed down in nEDM measurements, but future experiments are planned to increase the precision by up to two orders of magnitude~\cite{Ito:2018}.}
  \label{fig:edm}
\end{figure}
\noindent
\emph{Motivation}: The observation of permanent EDMs of protons, neutrons, nuclei,
atoms and molecules~\cite{Pospelov:2005pr,Chupp:2017rkp} would be deep probes of CP violation beyond the SM. Such violation of CP symmetry is necessary for baryogenesis~\cite{Sakharov:1967dj} and its discovery would be
central to understanding the origin of the baryonic matter in 
Universe~~\cite{Chupp:2017rkp}. The current limits of EDMs of most particles are 5--10
orders of magnitude larger than SM predictions for CP-violation arising from the CKM phase (Fig.~\ref{fig:edm}) . Since BSM physics will affect different systems differently, a combination of EDM tests in different systems is invaluable in constraining CP-violation scenarios.

LQCD is currently the best tool for connecting BSM physics to the EDMs of the nucleon, and to obtain the CP-odd pion-nucleon vertices that control the CP-odd nucleon-nucleon potential, which in turn determines nuclear EDMs. The current limit on the neutron EDM is $|d_n| < 3 \times 10^{-26} e\cdot {\rm cm}$  (90\% CL)~\cite{Baker:2006ts,Afach:2015sja}, which in turn can be used to estimate nuclear EDMs in EFTs and nuclear many-body models. In
the next decade, several neutron EDM experiments are planned (two of
which will be in the US, at the SNS and
LANL), and are expected to improve limits on the neutron EDM by
1--2 orders of magnitude, reaching a sensitivity of 
$\approx 5\cdot10^{-28}\,e\cdot\mathrm{cm}$.  The feasibility of a
storage-ring experiment to measure the proton EDM has been studied,
with a potential sensitivity of $10^{-29}e\cdot\mathrm{cm}$~\cite{Anastassopoulos:2015ura}, higher
than that expected for the neutron.  This technology may also lead to
measurements of EDMs of light nuclei ($^2$D, $^3$H, $^3$He). Experiments are also planned to improve the limits on EDMs of other heavy nuclei~(for a review of the experimental status and prospects see Ref.~\cite{Chupp:2017rkp}). Together, these
measurements will allow us to separate nuclear EDM contributions from
the intrinsic nucleon EDMs and nucleon CP-violating interactions.  The
isospin dependence of the intrinsic EDMs and nucleon CP-violating 
interactions will also be useful for understanding the pattern of CP violation in the BSM models~\cite{Cirigliano:2013lpa,Engel:2013lsa}.

\
\

\noindent
\emph{Progress report}: Although the discovery of an EDM in any of these experiments would be a breakthrough, constraining various BSM extensions will require a combination of 
different EDM measurements and matching from the nuclear level to
quark/gluon-level effective CP-violating operators, as discussed in the introduction.
Since the structure and interactions of the nucleon are described by 
QCD in the nonperturbative regime, chiral EFT extended to include BSM CP-violating operators (see 
e.g., \cite{Mereghetti:2010tp,deVries:2011an,deVries:2012ab,Bsaisou:2014oka} and references therein) along with LQCD calculations are needed to carry out this matching in a
model-independent way. At the quark and gluon level, there are several effective operators that may be
organized by their dimension.  
From the lowest-dimension dimension-4 QCD $\theta_{\rm QCD}$-term,\footnote{The $\theta_{\rm QCD}$ term is allowed in QCD as part of the SM, although its smallness is difficult to reconcile without extending SM.} to dimension-5(6) quark-EDM and quark/gluon chromo-EDM (cEDM) operators,\footnote{Low-energy dimension-5 operators arise from dimension-6 operators above the electroweak symmetry breaking scale. In many explicit BSM models, however, these operators are additionally
  suppressed due to the chirality violation that accompanies them and
  their contribution is no larger than the dimension-6 operators.} to the dimension-6 CP-violating 4-quark and 3-gluon interactions (Weinberg operator,
or gluon cEDM), these effective operators represent BSM CP-violating interactions
that are increasingly suppressed by the energy scale of the underlying new
physics. Quark/gluon CP-violating interactions can manifest themselves at the nuclear and atomic
level in two ways. First, they induce intrinsic EDMs in the proton and the neutron.  
Second, quark and gluon CP-violating interactions induce CP-violating nucleon and
nucleon-pion couplings that also contribute to nuclear EDMs.
Quantification of the nuclear EDMs, even for light nuclei, requires
low-energy nuclear effective theories and ab-initio nuclear many-body calculations that are based on nuclear EFTs~\cite{deVries:2011an,Bsaisou:2014oka}.  
For heavy nuclei, such as $^{199}$Hg and $^{255}$Ra, nuclear effects are expected to be significant (see Ref.~\cite{Engel:2013lsa} and references therein).  

\
\

\noindent
\emph{5-year goals and plans}: LQCD calculation of nucleon EDMs is similar to calculations of nucleon form factors, and its methodology is \emph{straightforward}.
However, the required statistical precision is more difficult to achieve
because of additional (CP-violating) interactions added as perturbations
to CP-even QCD interactions.
The difficulty may vary substantially depending on the CP-violating operator in
question, and in some cases the calculation may become \emph{challenging}
and require new methods for evaluating LQCD correlators.
The status of LQCD calculations with different CP-violating
operators is summarized below.
\begin{itemize}
\item[--]  
Calculations of isovector quark EDM-induced proton and neutron EDM ($p,n$EDM) are \emph{straightforward} since
they are given by the tensor charge up to electromagnetic corrections.  
Isoscalar quark EDM-induced $p,n$EDM calculations are also
\emph{straightforward}, even though the disconnected contributions (those arising from the interactions of sea quarks with the CP-violating currents) lead to more
noisy estimates. Relevant LQCD methodology has been substantially improved
in recent years, and first calculations for these operators already exist~\cite{Bhattacharya:2015esa}.
\item[--]  
Calculations of $\theta_\text{QCD}$-induced $p,n$EDMs~\cite{Shintani:2006xr,Shintani:2008nt,Shintani:2005xg,Berruto:2005hg,Aoki:2008gv,Guo:2015tla,Shindler:2015aqa,
Alexandrou:2015spa,Shintani:2015vsx} are procedurally straightforward, however the signal is elusive as
the expected value is small, i.e., it is $\propto m_q\propto m_\pi^2$ in chiral perturbation theory ($\chi$PT) at the leading order, where $m_q$ is the mass of the quark and $m_\pi$ is the mass of the pion, while the noise is large due to the fluctuations of the
topological charge that grow with the volume of the four-dimensional lattice.
Calculations with unphysical quark masses that lead to $m_\pi\gtrsim250$~MeV are underway
and their results (expected within 1--2 years) may be extrapolated to the
physical point using $\chi$PT.
However, calculations at the physical point, especially with quark actions that respect the chiral symmetry of QCD, will likely be \emph{challenging} and require more extensive resources.
\item[--]  
Calculations of isovector quark cEDM-induced $p,n$EDM are \emph{straightforward}, and
there are on-going efforts both with chiral and nonchiral quarks at the
physical point that show reliable signals~\cite{Abramczyk:2017oxr,Syritsyn:2018mon,Bhattacharya:2018qat}.  
Under renormalization, the isovector quark cEDM operator mixes
with the (isovector) pseudoscalar operator.
However, the latter can be ``rotated away'' by an isovector 
chiral transformation, with residual effects proportional to the isovector 
chiral-symmetry violations~\cite{Bhattacharya:2015rsa}.
\item[--]  
The isoscalar cEDM operator mixes with the isoscalar pseudoscalar quark operator, 
which is effectively the $\theta_\text{QCD}$ term, due to the axial anomaly.
Thus, LQCD determination of flavor-singlet cEDM-induced nucleon EDMs
depends on the \emph{challenging} calculations with $\theta_\text{QCD}$-term
described above.
In addition, the isoscalar quark cEDM operator will contribute to nucleon EDM
through numerically challenging quark-disconnected diagrams.
Therefore, even though calculations with the isoscalar quark cEDM are in many
respects similar to the isovector case, they are expected to be \emph{challenging} to obtain with comparable precision.
\item[--]  
Nucleon EDMs induced by the CP-violating 3-gluon operator (gluon cEDM) have been
studied in the quenched approximation~\cite{Dragos:2017wms}, and no
clear statistical signal has been observed as yet.
The study of this contribution is 
dependent on \emph{challenging} calculations with a $\theta_{QCD}$-term given the expected mixing.
A full calculation in QCD with physical light dynamical quarks should 
be considered at least \emph{challenging}.
\item[--]  
Calculations of the 4-quark CP-violating operator inducing $p,n$EDM have not been
attempted yet.
They will likely be very \emph{challenging} due to disconnected diagrams.
However, some special operators that are relevant to the phenomenology of specific
BSM models have only chirally-suppressed disconnected
contributions and mixing with lower dimensional operators.
Even though the calculation of these contributions is \emph{challenging}, there may be an opportunity to compute their contribution to $p,n$EDM with resources available within the next 5 years.
\item[--]  Each of the operators\footnote{The quark EDM only contributes at $\mathcal{O}(\alpha_{EM})$ and can be neglected.} discussed above can also give rise to CP violation in the $\pi NN$ and $NN\to NN$ interactions.  The former calculations can be done
  indirectly by studying the mass shift induced by the CP-even chiral partners of these operators~\cite{deVries:2015una,deVries:2016jox}, or by calculating the MEs of the axial current instead of the vector current as for the EDM. Most of these calculations are \emph{straightforward} or \emph{challenging} as above.
  \item[--]  A direct calculation of $N \to N\pi$ and any calculation of $NN\to NN$ MEs will be \emph{extremely challenging} because of the multiple hadrons in the initial/final state, requiring a simultaneous computation of the multi-level finite-volume spectrum and the use of L\"uscher and Lellouch-L\"uscher methodology to convert the LQCD output to elastic and inelastic transition amplitudes in these channels.
\end{itemize}

Finally, it must be noted that for all of the above calculations, quark-disconnected
contributions to the isoscalar vector current operator and hence to the EDM may
present an additional challenge.
While the present methodology for computing these contributions to CP-even
nucleon observables is efficient,
the presence of CP-violating interactions may enhance these contributions and
associated uncertainty.
Since no systematic study has been conducted to date (with the exception of quark EDMs),
such calculation may end up being \emph{challenging} instead of \emph{straightforward}.

\section{Baryon-number nonconservation and proton decay}
  \label{sec:pdecay}
\noindent
\emph{Motivation}: Baryon-number violation is one of Sakharov's conditions necessary for producing baryons at a different rate than antibaryons, hence accounting for the observed matter dominance in Universe~\cite{Sakharov:1967dj}. A well-searched-for experimental signature of baryon-number violation is the decay of the proton. Constraints on proton lifetime include $\tau_p > 8.2\times10^{33}$ years and $\tau_p > 1.4\times10^{34}$ years for the $p \to \pi^0e^+$ process reported in Refs.~\cite{Nishino:2009aa} and \cite{Babu:2013jba}, respectively, and $\tau_p > 5.9 \times 10^{33}$ years for the $p \to K+\nu^-$ process reported in Ref.~\cite{Abe:2014mwa}. Baryon-number conservation in the SM is not a direct consequence of fundamental symmetries such as gauge and Lorentz invariance, and is in fact broken nonperturbatively by the electroweak anomaly. Such violation, however, is too small to account for the observed baryon asymmetry in Universe, motivating baryon-number-violating (BNV) BSM scenarios at high scales. Given their enhanced gauge symmetry, Models of Grand Unified Theories~\cite{Pati:1973rp, Georgi:1974sy}, with or without supersymmetry, naturally predict that the proton can decay.

EFT provides a systematic framework for classification and analysis of BNV interactions that contribute to proton decay. At the lowest order, the BNV effective interactions can be written as dimension-6 three-quark--single-lepton operators inducing the decay of the proton to a pseudoscalar meson, such as pion, kaon and $\eta$, and an antilepton~\cite{Weinberg:1979sa, Wilczek:1979hc, Abbott:1980zj}. The effective Lagrangian consists of these interactions accompanied by the corresponding Wilson coefficients, computed perturbatively within a given high-scale model, and renormalized (customarily within the $\overline{MS}$ scheme) at $\mu = 2$ GeV where the corresponding operators can be renormalized nonperturbatively using LQCD~\cite{Aoki:2006ib, Aoki:2008ku, Aoki:2017puj}. This procedure provides a physical scale-independent cross section, which along with the experimental constraints on the rate, leads to constraints on the parameters of BNV models. Given the sensitivity of the cross section to the MEs of effective operators between the proton and the meson, and as next-generation experiments, such as DUNE and Hyper-Kamiokande, plan to improve the bounds on the proton decay rate significantly, it is important that these MEs are reliably estimated.

\
\

\begin{figure}[!t]
\centering
\includegraphics[width=1.0075\columnwidth]{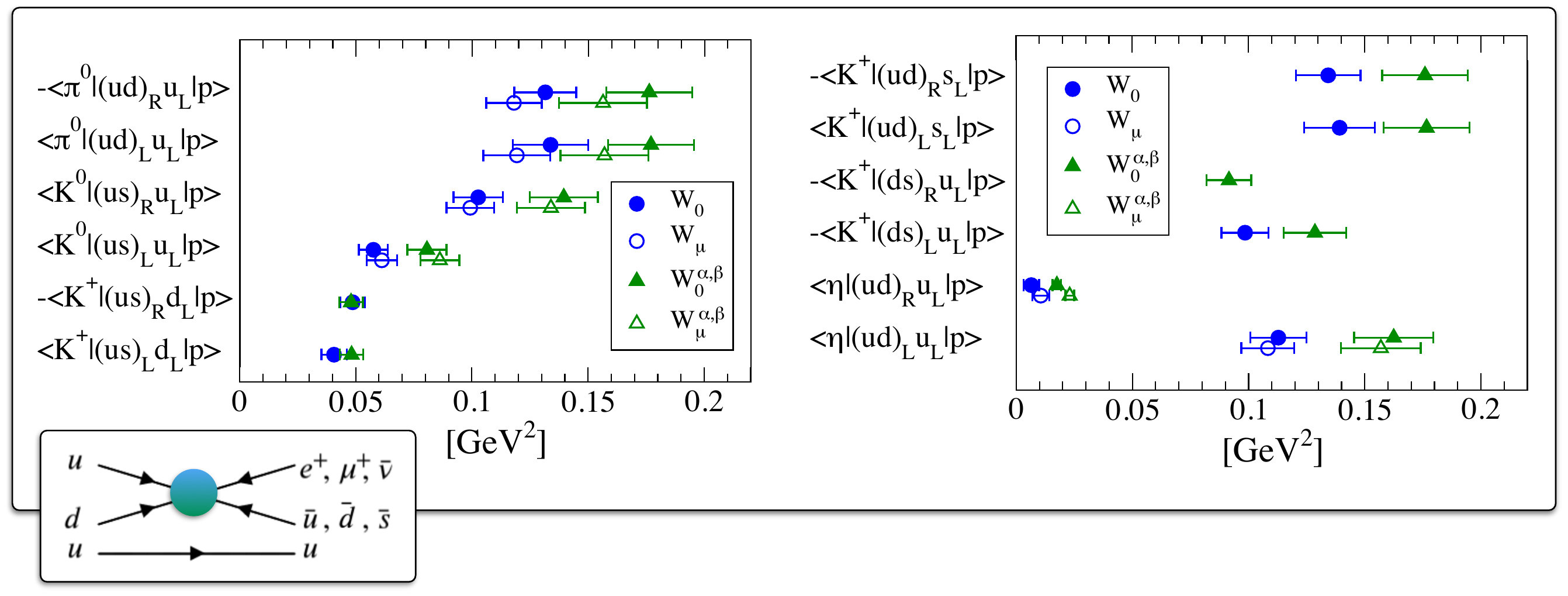}
\caption{The form factors parametrizing the shown MEs contributing to the proton decay at given values of momentum-transfer-squared, obtained from a recent LQCD study~\cite{Aoki:2017puj} using ``direct'' (blue) and ``indirect'' (green) methods with 2+1 flavor dynamical domain-wall fermions with Iwasaki gauge action generated by RBC and UKQCD collaborations~\cite{Aoki:2010dy}. $W_{\mu}$ is a given linear combination of $W_0$ and $W_1$ form factors, see Ref.~\cite{Aoki:2017puj}. The smaller panel is an illustration of the quark and lepton-level process mediated by an effective operator, denoted by the green circle.}
\label{fig:pdecay}
\end{figure}
\noindent
\emph{Progress report}: Since 1980s, LQCD has come a long way to compute the MEs relevant to proton decay~\cite{Aoki:1999tw, Tsutsui:2004qc, Aoki:2006ib, Aoki:2008ku, Braun:2008ur, Aoki:2013yxa, Aoki:2013yxa, Aoki:2017puj}. Uncontrolled systematic uncertainties in early calculations arising from the``quenched approximation'', chiral-symmetry violation, quark-mass extrapolation, and reliance on leading-order baryon $\chi$PT have been accounted for in state-of-the-art present-day calculations. From a computational perspective, a simpler computation is to evaluate the matrix elements of the proton-to-vacuum transition, which through baryon $\chi$PT can be related to two form factors, $W_{0}^{\alpha,\beta}$ and $W_{1}^{\alpha,\beta}$, characterizing the MEs contributing to proton decay. This is called the ``indirect'' method in literature~\cite{Claudson:1981gh, Aoki:1999tw}. Since at the physical kinematics the outgoing pion is far from the soft-pion limit, it is expected that the indirect method is systematically away from the result of a ``direct'' method, where the form factors $W_0$ and $W_1$ characterizing the ME for proton to pion (meson) are computed directly. The most recent work reports on results of a LQCD study with 2+1 dynamical flavor domain wall fermions at several values of the light quark masses~\cite{Aoki:2017puj}. Fig.~\ref{fig:pdecay} presents the values of relevant form factors for corresponding MEs (normalized in the $\overline{MS}$ scheme at $\mu = 2$ GeV) as a function of momentum-transfer squared in the process, using both the direct and indirect methods, at larger-than-physical values of quark masses, but with the associated systematics taken into account.

\
\

\noindent
\emph{5-year goals and plans}: The calculation of the MEs relevant to the proton decay in the direct method at the physical values of the quark masses is an ongoing effort within the USQCD collaboration and elsewhere, see e.g., Ref.~\cite{Yoo:2018fyn}. With new error-reduction techniques, such as all-mode-averaging~\cite{Blum:2012uh, Blum:2012my, Shintani:2014vja}, these calculations are considered \emph{straightforward}. Similarly, while complete studies including continuum and volume extrapolations are computationally challenging, they will remian procedurally \emph{straightforward}, and hence a target precision of $< 10\%$ is within the reach in the next 5 year.

Other plausible final states for the decay of the proton are two-meson--single-antilepton states. In fact, model estimations point to a higher branching ratio for $p \to \pi \pi e^+$ in the isospin-singlet channel compared with the $p \to \pi e^+$ channel~\cite{Wise:1980ch}. As the upcoming DUNE will be sensitive to such three-body final-state processes, it is important that the corresponding hadronic matrix elements will be computed reliably with LQCD. The two-hadron final state complicates the analysis of the finite-volume matrix elements, hence making these calculations more \emph{challenging} than those already performed for the single-hadron final states. However, the formalism and methodologies are known~\cite{Lellouch:2000pv,Detmold:2004qn,Briceno:2012yi,Agadjanov:2014kha,Briceno:2014uqa,Briceno:2015csa,Briceno:2015tza}, and are successfully implemented in similar processes such as $K \to \pi \pi$ decay~\cite{Bai:2015nea} and the $\rho \to \pi \gamma^*$ transition~\cite{Briceno:2016kkp}, such that the extraction of the physical matrix elements from LQCD calculations at arbitrary kinematics will not be out of reach in upcoming years.

Finally, a question that remains to be investigated is the effect of nuclear medium, as is the case in experiments, on the decay rate of the proton. Addressing this question requires a study of the matrix elements relevant for transition $NN \to NP$, where $N$ and $P$ denote the nucleon and the pseudoscalar meson, respectively. Estimating the size of two-nucleon short-distance contribution in many-nucleon systems with the help of nuclear EFTs will be the next step. The LQCD calculation of the matrix element with two-nucleon initial and/or final state will be \emph{challenging} at the physical values of the quark masses, but are technically similar to those studied in Refs.~\cite{Savage:2016kon,Shanahan:2017bgi,Tiburzi:2017iux,Chang:2017eiq} in the context of MEs of scalar, axial and tensor currents in light nuclei, see Secs.~\ref{sec:0vbb} and \ref{sec:darkmatter}.

\section{Lepton-number nonconservation and neutrinoless double-$\beta$ decay of a nucleus}
  \label{sec:0vbb}
\noindent
\emph{Motivation}: The observed matter-antimatter asymmetry in the visible Universe requires the violation of baryon-number conservation. As the structure of the SM ensures that fluctuations in the number of baryons and leptons are equal, observing the nonconservation of lepton number through the \emph{$0\nu\beta\beta$ decay} of nuclei would provide a direct probe of baryon-number nonconservation, a critical process in the evolution of Universe. Its observation would shed light on the nature of neutrinos and will unambiguously prove that the neutrino is a Majorana fermion, \emph{i.e.}, it is its own antiparticle. Although solely observing a decay will mark a major discovery, without a proper isolation of the nuclear MEs contributing to the rate of this decay, little insight can be gained into the BSM mechanism mediating this transition. Furthermore, the design and implementation of the DOE's new tonne-scale experiment, that has been recommended with high priority in the 2015 Long Range Plan for Nuclear Sciences~\cite{Geesaman:2015fha}, strongly relies on more accurate estimations for the rate of this process in target nuclei. Reduced uncertainties in theoretical calculations of nuclear MEs with \emph{ab initio} methods are highly desired, and are underway~\cite{Engel:2016xgb}.

\
\

\noindent
\emph{Progress report}: The LQCD community, including members of the USQCD collaboration, has engaged in the larger $0\nu\beta\beta$-decay program~\cite{Nicholson:2018mwc,Shanahan:2017bgi,Tiburzi:2017iux}. The goal of LQCD calculations is to reach the level of precision needed for the few-nucleon MEs to be valuable to the nuclear many-body physicists who aim to constrain the rate of this process in heavier nuclei. Fast progress is being made in developing a path from the QCD input for few-body MEs to nuclear many-body calculations, and is built upon applicable nuclear EFTs~\cite{Prezeau:2003xn,Menendez:2011qq,Cirigliano:2017tvr,Cirigliano:2018hja,Cirigliano:2018yza,Pastore:2017ofx}, as illustrated in Fig.~\ref{fig:0vbbchart}. Two types of MEs will be the focus of LQCD studies: \emph{i)} MEs of nonlocal SM quark bilinear operators that change the total isospin of two-nucleon systems by two, $\Delta I=2$, and include the exchange of a light Majorana neutrino between the two operator insertions. Such a scenario arises from a minimal extension of the SM. ii) MEs of local non-SM four-quark--two-lepton operators that shift the lepton number $L$ and the isospin of the initial state by two. These arise from integrating out new particles  with masses larger than the EW scale,  see e.g. Fig.~\ref{fig:0vbbchart}. It is important to note that both of these possibilities may be equally important, each with their own signatures for the experimental observables, and must be accounted for when making predictions for potential observations.

\begin{figure}[!t]
\centering
\includegraphics[width=1\columnwidth]{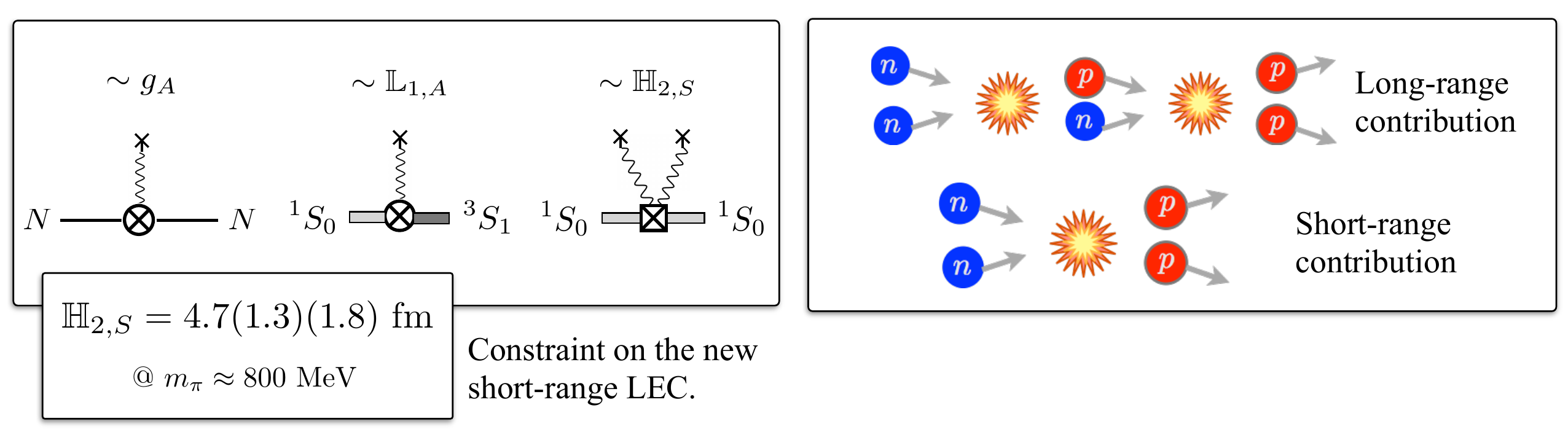}
\caption{As a preliminary step towards constraining the MEs relevant to $0\nu\beta\beta$ with a light Majorana neutrino, the NPLQCD collaboration performed a calculation of the ME of two axial vector currents that induce the SM counterpart of $0\nu\beta\beta$ decay process, namely the $2\nu\beta\beta$ decay, in the two-nucleon system~\cite{Shanahan:2017bgi,Tiburzi:2017iux}. This first study was performed at a single lattice spacing and lattice volume and with quark masses corresponding to a pion mass of $\approx800~\text{MeV}$, at which point the two-nucleon systems are found to be rather deeply bound~\cite{Beane:2012vq}. The new short-range $\Delta I=2$ operator in the two-nucleon EFT was identified and its corresponding LEC, $\mathbb{H}_{2,S}$, was constrained from the result of the LQCD calculation. This study will be further developed in the next five years to account for the exotic process of $0\nu\beta\beta$ decay, and towards physical values of quark masses, such that it will impact \emph{ab initio} nuclear structure calculations of this process in larger nuclei.}
\label{fig:2vbbnplqcd}
\end{figure}

Considering the first scenario, a closely related SM process in the two-nucleon sector is $nn \to ppee\overline{\nu}\overline{\nu}$. This process occurs at very low energies and hence is well-suited for constraining the leading $\Delta I=2$ interaction in a nuclear EFT. Although the neutrinoless process involves radically different kinematics given the presence of a virtual Majorana neutrino, constraining the nuclear MEs relevant for $2\nu\beta\beta$ decay can provide much insight into the importance of new short-distance LECs, the size of multi-body currents, and the significance of the ``quenching'' of the axial charge in isotopes that undergo $\beta\beta$ decay. USQCD's NPLQCD collaboration took the first steps towards constraining the MEs relevant for this process using ensembles of gauge-field configurations with a single lattice spacing, a single lattice volume and larger-than-physical values of quark masses, corresponding to a pion mass of $\approx800~\text{MeV}$~\cite{Shanahan:2017bgi,Tiburzi:2017iux}. This calculation led to a constraint on the contact LEC in the two-nucleon sector that isolates the contributions from high-energy intermediate states beyond an on-shell deuteron, see Fig.~\ref{fig:2vbbnplqcd}. The natural next step in this program is to recompute this ME with the inclusion of the Majorana neutrino, which is an ongoing project within the USQCD collaboration~\cite{Detmold:2018zan} and elsewhere~\cite{Feng:2018pdq}.

\begin{figure}[!t]
\centering
\includegraphics[width=0.9\columnwidth]{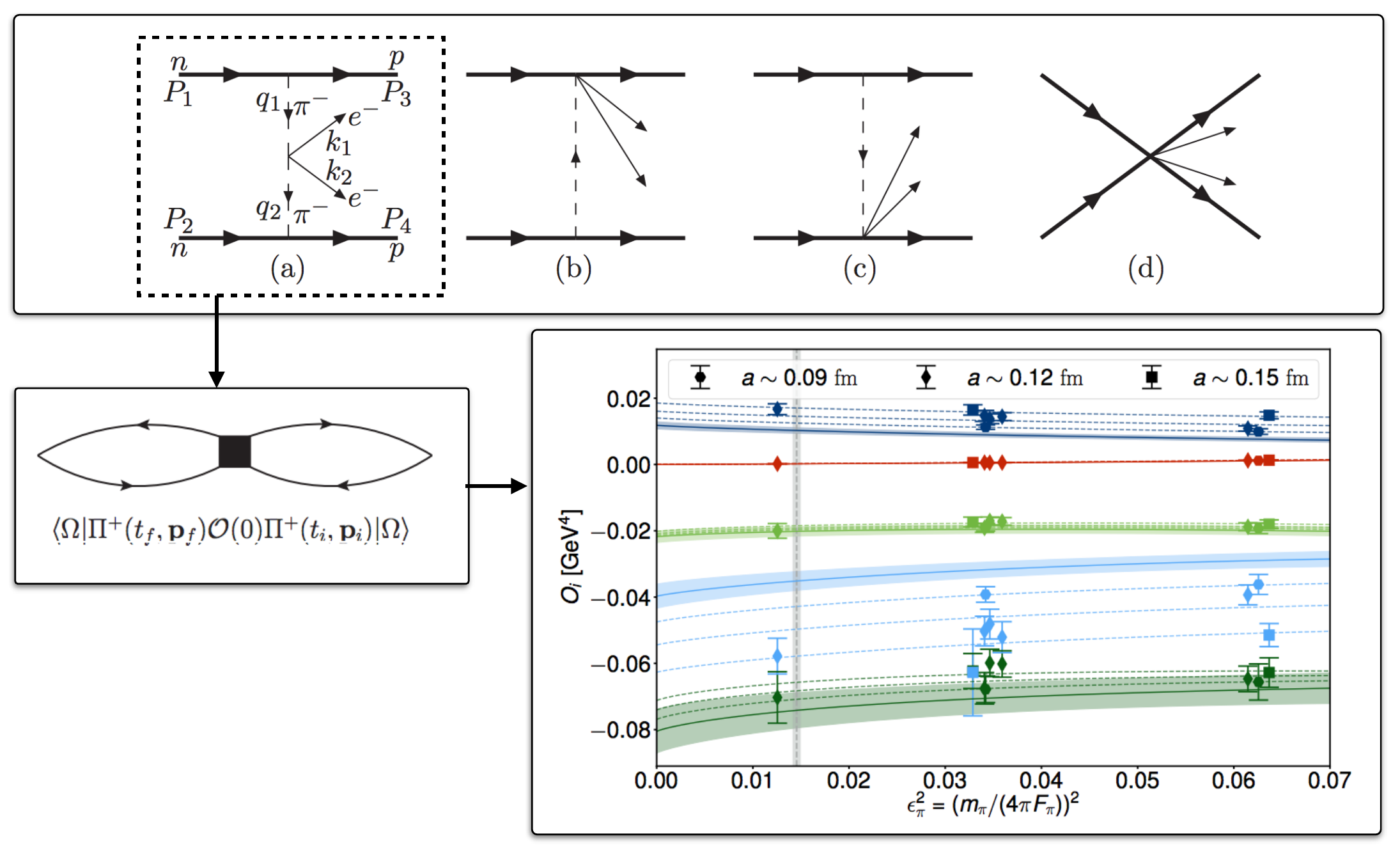}
\caption{Within the Weinberg power counting of nuclear forces, an organizational scheme was proposed in Ref.~\cite{Prezeau:2003xn} to identify the dominant contributions to the $0\nu\beta\beta$ decay of two neutrons at low energies, as depicted in the upper panel. The CalLat collaboration performed the first calculation of the MEs of $\Delta L=2$ four-quark operators (arising from heavy-scale scenarios for $0\nu\beta\beta$) in the pion, hence constraining part of the amplitude for $nn \to ppee$ in which the pion exchanged between the two nucleons undergoes a $0\nu\beta\beta$ decay~\cite{Nicholson:2018mwc}. The study was performed on USQCD's MILC ensembles of gauge-field configurations and included continuum, finite-volume and chiral extrapolations. The lower-right plot displays the MEs $ \langle  \pi^+ | O_i   | \pi^- \rangle$ corresponding to a set of $\Delta I =2$ four-quark operators, $O_i$, as defined in the reference. Within the next five years, further developments will enable computations of more challenging $\langle p \pi^+ | O_i |n \rangle$,  $\langle pp | O_i | nn \rangle$ MEs.}
\label{fig:0vbbcallat}
\end{figure}

Considering the second scenario, if the Weinberg power counting of nuclear forces is applied, one would conclude that the dominant contribution to the amplitude of $nn \to ppee$ arise from a process in which the exchanged pion (responsible for the long-range piece of the nuclear forces) undergoes a $0\nu\beta\beta$ decay, as demonstrated in the upper panel of Fig.~\ref{fig:0vbbcallat}. The CalLat collaboration, including members of the USQCD collaboration, has taken the first initiative to compute the MEs relevant to short-distance scenarios~\cite{Nicholson:2018mwc}. The MEs of several dimension-9 $\Delta I=2$ four-quark operators have been evaluated in the pion, and have been extrapolated to the continuum, infinite-volume and physical masses of quarks using USQCD's MILC gauge-field ensembles.\footnote{Similar results to the ones in Ref. [42], although with larger uncertainties, were obtained in Ref.~\cite{Cirigliano:2017ymo} by relating the $\pi^- \to \pi^+$  MEs via chiral $SU(3)$ symmetry to $K^0$-$\overline{K}^0$ MEs known from LQCD~\cite{Bertone:2012cu,Boyle:2012qb,Jang:2015sla}.} As will be discussed below, the more challenging MEs that involve nucleonic final states will be studied in upcoming years such that both the EFT and the assumed power-counting scheme can be examined.
   
\
\

\noindent
\emph{5-year goals and plans}: In a model-independent approach, we organize the discussion of LQCD contributions that will be computed in the next five years according to the underlying mechanism (\emph{i.e.}, dimension of the $\Delta L=2$ operator), as well as their level of difficulty. Within an EFT approach to Lepton Number Violation (LNV), operators with $\Delta L=2$ arise at odd dimensions  starting at dimension five~\cite{Weinberg:1979sa, Lehman:2014jma,Graesser:2016bpz,
Kobach:2016ami}. As was alluded to above, depending on the scale of new physics and the mechanism by which it appears, operators of dimension five, seven and nine can contribute to $0\nu\beta\beta$ decay at levels comparable to the current and planned experimental sensitivities:
\begin{itemize}
\item[--]  LNV from the dimension-5  operator (light Majorana-neutrino exchange): Here, the main phenomenological goal is to assess what kind of sensitivity the next generation searches will have to the effective neutrino Majorana mass, $m_{\beta \beta} = \sum_i  U_{ei}^2  m_i$. To connect to the nuclear ME in larger nuclei, a LQCD calculation must first determine the MEs  $ \langle  \pi^+ | S_{NL} | \pi^- \rangle$, $ \langle  p \pi^+ | S_{NL} | n  \rangle$,  $ \langle pp | S_{NL} | nn \rangle$, where the nonlocal effective action $S_{NL}$ (up to factors of $G_F$,  $m_{\beta \beta}$, etc.) is defined as follows:
\begin{eqnarray}
S_{NL} =  \int  d^4x \, d^4y   \, S_0  (x-y)  \  T \left\{  J^+_{\alpha}  (x)  J^+_{\beta} (y) \right\}  \, g^{\alpha \beta}~.
\end{eqnarray}
Here, $S_0(x-y)$ is the massless scalar propagator representing the Majorana neutrino propagator and $J_\alpha^+ =  \bar{u} \gamma_\alpha (1 - \gamma_5) d$, where $u$ and $d$ are up and down quarks. The $nn \to pp$ ME is particularly important given that  a consistent chiral EFT approach requires a leading-order contact interaction~\cite{Cirigliano:2018hja}, whose coupling can be reliably determined only by matching to a LQCD calculation in the two-nucleon sector. The calculations that are planned for the next five years will all require new developments and can be categorized as: \emph{ii) Challenging}: The $\pi^- \to \pi^+$ ME offers a warm-up calculation, since it is simpler to implement and its contributions comes at higher orders in an EFT matching. The required lattice methodology has overlap with the calculation of electromagnetic corrections to hadronic and semi-leptonic processes, rare meson decays such as $K^+ \to \pi^+ \nu \overline{\nu}$, and the light-by-light contribution to the muon $g-2$. Ongoing effort within the USQCD collaboration is focused on this ME and preliminary results at a range of quark masses will become available in the next 1-2  years. \emph{iii) Extremely challenging}: While the lattice methodology developed for the pionic ME can be applied to $ \langle  p \pi^+ | S_{NL} | n  \rangle$,  $ \langle pp | S_{NL} | nn \rangle$, the signal-to-noise degradation in nuclear systems will demand far more computational resources to obtain the first results at the physical point. Further, complexities arising from two-particle initial and/or final states must be dealt with through new formalisms, that are yet to be developed, to connect the physical ME of a nonlocal operator to that obtained in a finite-volume Euclidean LQCD calculation. In the next five years, such a formalism will be in place and  calculations of the new short-distance LECs with $\sim 20\%-30\%$ uncertainty will be plausible with a range of quark masses, such that the first extrapolation to the physical point can be achieved. The projected precision may therefore be limited due to extrapolation uncertainties. The expected uncertainty will depend largely on the availability of ensembles with sufficiently large volumes and a range of lighter quark masses, which assuming the same trend in availability of computational resources to the USQCD program, are likely to become a reality within 5 years.
\item[--] LNV from dimension-9 operators (``short-distance" mechanisms): Here, the main 
phenomenological goal is to assess what kind of models (beyond the high-scale seesaw mechanism) 
can produce $0\nu\beta\beta$ decay at levels that can be observed in ongoing searches. Improving the hadronic uncertainties will be crucial in comparing the reach of $0\nu\beta\beta$ decay and LNV signatures at colliders, such as in the $pp \to e e + 2$~jets process~\cite{Peng:2015haa}. At low energies of the order of $\sim$ 1 GeV, the operator basis involves a handful of six-fermion local operators, which are factorized as lepton bilinears times charge-changing four-quark operators, $O_i$~\cite{Prezeau:2003xn,Graesser:2016bpz,Cirigliano:2018yza}. LQCD can provide the MEs of the four-quark charge-changing local operators between the appropriate 
hadronic states, namely  $ \langle  \pi^+ | O_i   | \pi^- \rangle$, $ \langle  p \pi^+ | O_i   |n     \rangle$,  $ \langle pp | O_i   | nn \rangle$. The calculations that are planned for the next five years can be categorized as: \emph{ii) Challenging}: While the computation of $ \langle  \pi^+ | O_i   | \pi^- \rangle$ MEs have been recently completed by the CalLat collaboration~\cite{Nicholson:2018mwc}, one recalls that in a consistent chiral power counting~\cite{Cirigliano:2018yza}, input from  $ \langle pp | O_i   | nn \rangle$ may be as important as the pionic contribution. As discussed above, such calculations are more challenging due to the signal-to-noise degradation. However, both the lattice technology for computing local MEs and the finite-volume formalism that enables the extraction of corresponding physical MEs are already in place, making the implementation of these calculations plausible in the next five years. Once again, the availability of gauge-field ensembles suitable for studying nuclear systems with large characteristic length scales is a key to the success of this program.
\end{itemize}
Finally, we must emphasize that given the lack of experimental input on the short-distance contributions to the nuclear MEs in both the light Majorana exchange and the short-distance scenarios,  the input from LQCD calculations is crucial in advancing the theoretical frontier of the $0\nu\beta\beta$ program. This also means that even a constraint on these MEs and/or EFT couplings at the level of a few tens of percent will be extremely useful in enhancing the ongoing and upcoming \emph{ab initio} nuclear structure calculations that may be missing large contributions from currently unknown short-range effects.

  \section{Baryon-number minus lepton-number nonconservation and $n \mhyphen \bar{n}$ oscillation}
  \label{sec:nnbar}
\noindent
\emph{Motivation:} Baryon number is approximately conserved to a very high precision in the SM at low energies, but only baryon minus lepton number (B-L) is an exact symmetry of the SM.
Baryogenesis mechanisms generically require B-L violation since B-violating electroweak sphalerons will washout any high-scale baryon asymmetry generated in the early Universe unless there is a B-L asymmetry.
Low-scale B-L violation and post-sphaleron baryogenesis can be compatible with present experimental limits on B and B-L violation if they do not lead to proton decay (this could be because of selection rules imposed by the mechanism of baryon-number violation) and instead proceed through higher-dimensional operators and lead for instance to neutron-antineutron oscillations.
Experimental searches for B-L neutron-antineutron ($n \mhyphen \bar{n}$) oscillations can test some of these low-scale baryogensis scenarios and provide complementary searches to B violation in proton decay and B-L violation through L violation in $0\nu\beta\beta$ decay, see e.g., Refs.~\cite{Mohapatra2009,Babu2013,Phillips2016} for reviews and further references.
 
Two general classes of experiments can be used to search for $n \mhyphen \bar{n}$ oscillations: slow neutron beam experiments and detection of $n \mhyphen \bar{n}$ annihilation in nuclei.
Neutron beam experiments are advantageous because of their clean theoretical interpretation. The tightest constraints from cold neutron beam experiments on the $n \mhyphen \bar{n}$ oscillation time $\tau_{n \mhyphen \bar{n}}$ were obtained at the Institut Laue-Langevin (ILL) and give $\tau_{n \mhyphen \bar{n}} > 0.86\times 10^8$ s~\cite{Baldo-Ceolin1994}.
Future experiments have been proposed at the European Spallation Source (ESS) ~\cite{Milstead2015,Frost2016} that could improve bounds on $\tau_{n \mhyphen \bar{n}}$ by a factor of 32, as well as at other reactors~\cite{Fomin2018}.
Nuclear $n \mhyphen \bar{n}$ annihilation searches in large underground volume detectors provide competitive or even stronger bounds on B-L violation than neutron beam experiments, but are more difficult to interpret theoretically.
Super-K provides a bound on the lifetime of Oxygen-16 of $\tau_{^{16}\text{O}} > 1.9\times 10^{32}$ years~\cite{Abe2015}, which nuclear structure calculations indicate is associated with a constraint on the vacuum oscillation time of $\tau_{n \mhyphen \bar{n}} > 2.7\times 10^8$ s~\cite{Friedman2008}.
Bound on the $n \mhyphen \bar{n}$ transition time in the deuteron, $\tau_{D} > 1.18\times 10^{31}$ years~\cite{Aharmim2017}, also provides a competitive bound on the vacuum oscillation time, $\tau_{n \mhyphen \bar{n}} > 1.6 \times 10^8$ s~\cite{Oosterhof:2019dlo}.
Future underground neutrino facilities such as DUNE may provide stronger limits on B-L violation in nuclei~\cite{Hewes2017}.

\
\

\begin{figure}[t]
  \centering
  \includegraphics[width=\textwidth]{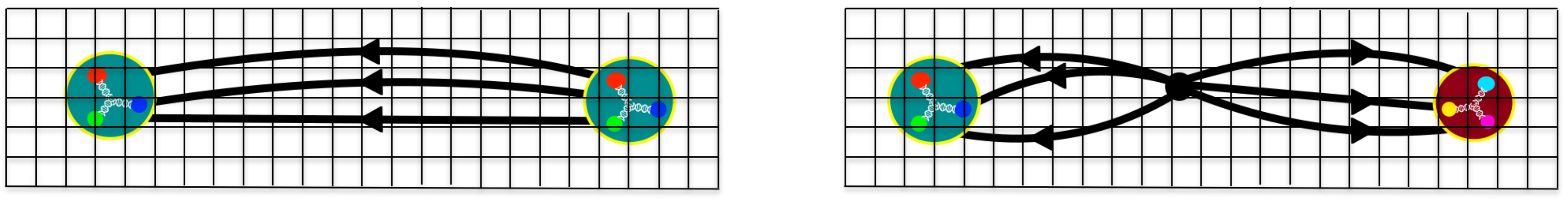}
  \caption{The left diagram schematically shows how a neutron two-point correlation function can be constructed from three copies of a single quark propagator. Calculations of three-point functions including operator insertions are typically more expensive because they require multiple quark propagator calculations, but using the construction schematically depicted on the right, $n \mhyphen \bar{n}$ MEs can be constructed from the same quark propagator used to construct the neutron two-point function. The quark propagator source point is taken to be the operator insertion point (black dot) and extends forward in time to a neutron annihilation operator or backward in time to an anitneutron creation operator.
    \label{fig:nnbar}}
\end{figure}

\noindent
\emph{Progress report:} The results of present and future searches for $n \mhyphen \bar{n}$ oscillations and associated B-L violation in nuclei can be interpreted in an EFT framework where BSM effects are encoded in non-zero LECs for dimension-9 six-quark operators.
A LQCD calculation of the $n \mhyphen \bar{n}$ transition MEs for a complete basis of six-quark operators~\cite{Chang:1980ey,Kuo1980,Rao1984} can be used to constrain BSM theories of B-L violation with experimental results for $\tau_{n \mhyphen \bar{n}}$ and improve upon estimates from the MIT bag model~\cite{Rao1984}.
Exploratory calculations involving USQCD members of such a complete basis of six-quark operators were performed using an efficient operator construction shown in Fig.~\ref{fig:nnbar}~\cite{Buchoff2012}.
The perturbative and non-perturbative renormalization of $n \mhyphen \bar{n}$ transition operators was subsequently studied in a chiral basis that greatly simplifies $n \mhyphen \bar{n}$ operator renormalization~\cite{Syritsyn2015,Buchoff2016}.
Non-perturbative RI/MOM renormalization factors were calculated for the $n \mhyphen \bar{n}$ operators in Refs.~\cite{Syritsyn2015,Rinaldi:2018osy,Rinaldi:2019thf} using a LQCD ensemble with approximate chiral symmetry and physical quark masses generated by the RBC/UKQCD collaborations~\cite{Blum:2014tka}.
The same ensembles were used for a recent calculation of the renormalized $n \mhyphen \bar{n}$ transition MEs in Ref.~\cite{Rinaldi:2018osy} involving USQCD members.
LQCD results suggest that MEs are about $5-10$ times larger than had previously been estimated using MIT bag model results and therefore that the reach of current and future experiments into BSM parameter space is larger than expected.

\
\

\noindent
\emph{5-year goals and plans:}  Future calculations of the six-quark MEs studied in Ref.~\cite{Rinaldi:2018osy} are \emph{straightforward} and could reduce the statistical uncertainties in current results as well as discretization effects and other systematic uncertainties.
Higher dimensional six-quark operators including electromagnetic current insertions have also been identified as candidates for B-L violating neutron-antineutron conversions not suppressed by the presence of magnetic fields~\cite{Gardner:2017szu}.
LQCD MEs of these operators could be computed using similar techniques as existing six-quark operator calculations and should be considered \emph{straightforward}.

Allowing nuclear decay experiments to be interpreted with the same level of theoretical rigor requires \emph{challenging} calculations of six-quark MEs in light nuclei. Chiral EFT has been used recently to calculate the deuteron lifetime $\tau_D$ in terms of $\tau_{n\bar{n}}$~\cite{Oosterhof:2019dlo} and therefore allows robust constraints on BSM theories to be extracted from a combination of SNO constraints on $\tau_D$~\cite{Friedman2008}, chiral EFT and LQCD. Direct LQCD calculations of B-L violating deuteron annihilation could be used to test chiral EFT for this process and constrain higher-order LECs. Calculations in other light nuclei could be used to test and inform nuclear many-body models needed to predict the lifetimes of larger nuclei such as $^{16}$O in Super-K and $^{40}$Ar at DUNE. At unphysically large values of the quark masses these calculations may require a similar resource investment as nuclear MEs for other SM and BSM currents performed by the NPLQCD collaboration~\cite{Savage:2016kon,Shanahan:2017bgi,Tiburzi:2017iux,Chang:2017eiq}.
Performing these calculations at physical values of the quark masses would be \emph{extremely challenging} because of the complex few-body final states and would require new formal developments as well as high statistics.

  \section{Lepton-flavor nonconservation and muon to electron conversion}
  \label{sec:mue}
\begin{figure}[!t]
\centering
\includegraphics[width=0.9\columnwidth]{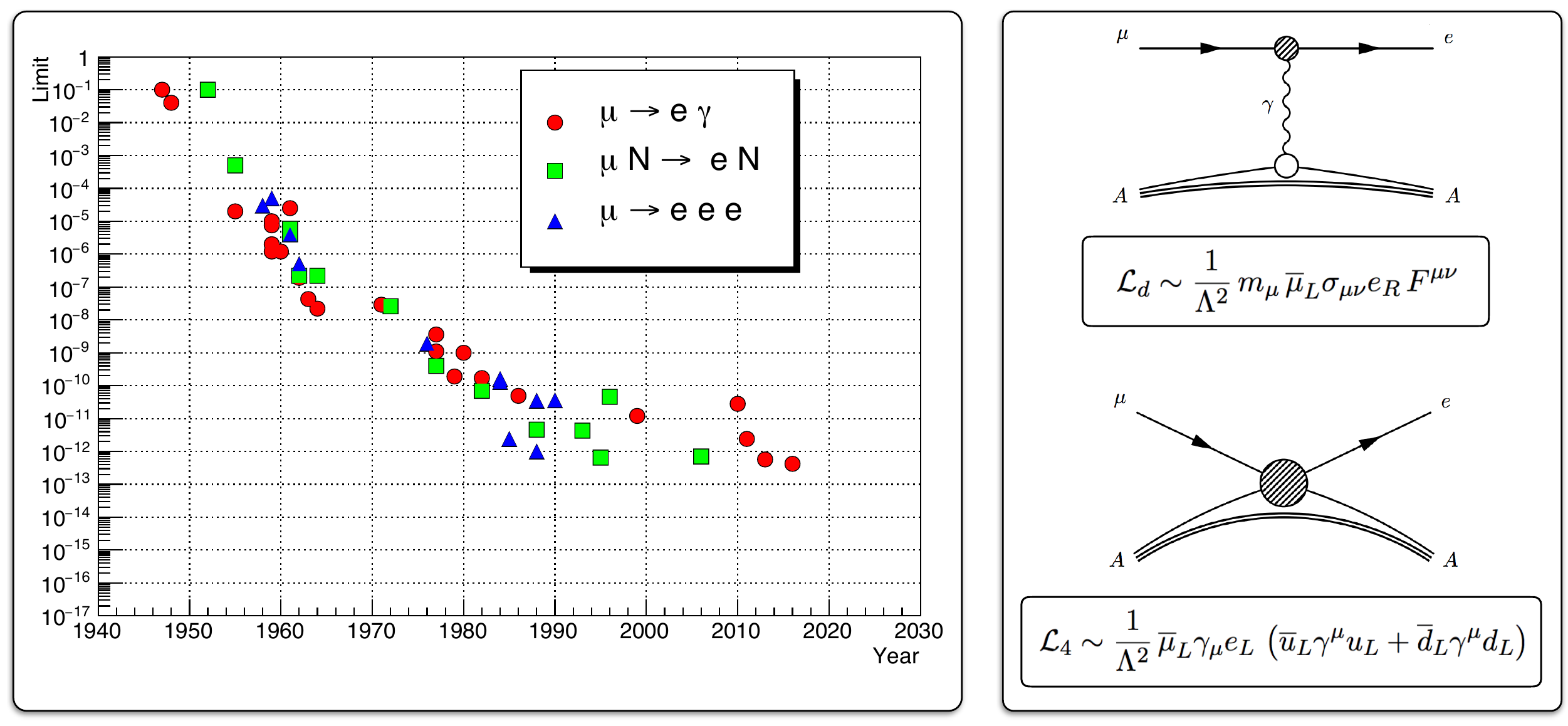}
\caption{The left panel shows limits on the branching ratio of LFV muon decays as a function of the year~\cite{Calibbi:2017uvl}. The right panel depicts examples of LFV interactions induced by the dimension-6 operators. $A$ denotes the atomic number of the target nucleus.}
\label{fig:mu2e}
\end{figure}

\noindent
\emph{Motivation}: The conservation of lepton family number  is an accidental symmetry of the SM with no right-handed neutrinos, 
i.e., it follows simply from the field content, gauge symmetry, and  the inclusion of operators of dimension less than or equal to four.
With the discovery of neutrino oscillations, we now have evidence that lepton flavor is not conserved.  However, 
in minimal extensions of the SM that include only neutrino masses and mixing, branching ratios (BRs) for lepton-flavor-violating (LFV) decays of charged leptons are small ($<10^{-54}$) due to a quadratic Glashow-Iliopoulos-Maiani (GIM) suppression mechanism~\cite{Petcov:1976ff,Marciano:1977wx} and are likely unobservable. 
Conversely, assuming new LFV dynamics at a scale $\Lambda_{\text{LFV}}$, LFV processes such as $\mu \to e \gamma$,  $\mu \to 3 e$, 
and  $\mu \to  e$ conversion in nuclei probe scales up to $10^3$~TeV at the current and future experimental sensitivity level (BR $\sim 10^{-13} - 10^{-16}$). 
It is anticipated that the experimental sensitivity in $\mu \to  e$ conversion in nuclei will be greatly enhanced through the Mu2e experiment 
at Fermilab~\cite{Giovannella:2018dly} and the COMET experiment in J-PARC~\cite{Litchfield:2018gez}. The left panel of Fig.~\ref{fig:mu2e} shows the limits on the branching ratio of several LFV decays of the muon over time.
  
In order to understand the implications of current and future searches 
of  $\mu \to  e$ conversion in terms of underlying  LFV mechanisms, 
one needs to control the associated hadronic/nuclear uncertainties.   
In fact,  explicit studies have shown how	reliable MEs will help establish the  pattern of LFV signatures in various decay channels depending on the underlying 
mechanism~\cite{Cirigliano:2009bz,Crivellin:2014cta}.
Particularly interesting from a phenomenological standpoint is Higgs-mediated lepton-flavor violation,  which depends on possible non-standard couplings 
of the 125 GeV Higgs boson and on possibly richer Higgs sectors. 
The phenomenological analysis in any underlying model is facilitated  by an EFT setup~\cite{Cirigliano:2009bz,Cirigliano:2017azj,Crivellin:2017rmk}.
In this context, one needs MEs of various quark billinears in nuclei of experimental interest, e.g., $^{27}$Al for next generation. Examples of such LFV interactions arising from dimension-6 operators are depicted in the right panel of Fig.~\ref{fig:mu2e}.

\
\

\noindent
\emph{Progress report}: LQCD input to the LFV program is closely related to that required in the dark-matter detection program. The progress in obtaining these MEs in the zero and nonzero-recoil limit using LQCD is mentioned in Sec.~\ref{sec:darkmatter}.

\
\

\noindent
\emph{5-year goals and plans}: The analysis of a muon to electron conversion process in a nucleus can be organized using chiral EFT, according to which  the leading term in the $\mu \to e$ transition amplitude is controlled by the coupling of the external probe (quark density) to single nucleons. Therefore, the needed LQCD input at this stage is the set of nucleon form factors (scalar, vector, axial, tensor and pseudoscalar) at $q^2= m_\mu^2$, where $q$ is the momentum transfer to nucleon and $m_{\mu}$ is the muon mass.
Most relevant for Higgs-mediated processes is the set of scalar form factors (with  $u,d,s$ flavor) and the MEs of the gluonic ($GG$) operator (that is induced by integrating out heavy quarks).  
For spin-dependent mechanisms, other tensor structures, especially the flavor diagonal axial form factor, becomes relevant. Since one needs MEs of flavor-diagonal  quark bilinears, 
the calculation of disconnected diagrams is necessary. 
This set of calculations  would at first sight be considered as \emph{straightforward}.  However, in the case of the scalar densities, it may be \emph{challenging} to reach a few-percent precision in order to compete with non-LQCD determinations of the sigma term~\cite{Hoferichter:2015dsa}. The reason lies in the need to evaluate computationally expensive quark disconnected contributions, for which considerable progress has been made by the collaboration in recent years, as detailed in a companion whitepaper~\cite{Joo:2018qcd}. For the form-factor calculations from LQCD, either very large volumes ($\gtrsim 12$ fm) or modified boundary conditions~\cite{Bedaque:2004kc,deDivitiis:2004kq} are required to access the low-momentum transfers relevant to the $\mu \to e$ process, which will be computationally \emph{challenging} but procedurally straightforward.

Beyond the leading chiral order, two-nucleon contributions to the $\mu \to e$ conversion in a nucleus arise. 
Existing estimates show that effect is non-negligible for scalar interactions in $^{27}$Al~\cite{Bartolotta:2017mff}. 
LQCD can provide useful input in this regard,  by computing directly the appropriate MEs in light nuclei~\cite{Chang:2017eiq}. 
 The connection to  nuclei of experimental interest would be achieved by matching the LQCD result 
 to a chiral EFT calculation in the few-body systems and use the resulting operators in heavier nuclei.   
 This  set of calculations is \emph{challenging} to \emph{extremely challenging}.  
Going beyond the leading chiral order is well motivated in the case of Higgs-mediated lepton-flavor violation (scalar operators), 
as future $\mu \to e$ conversion searches provide the best probe of these scenarios.

   \section{Dark-matter cross sections with nucleon and nuclei}
  \label{sec:darkmatter}
\noindent
\emph{Motivation:} There is abundant evidence from galactic rotation curves, gravitational lensing, structure formation, and the cosmic microwave background that dark matter (DM) makes up a large percentage of the matter in Universe~\cite{Bertone:2004pz,Bertone:2016nfn}. Many models provide weakly interacting massive particles (WIMPs) as dark matter candidates, e.g., neutralinos in supersymmetric extensions of the SM. Interpreting direct dark matter searches in the context of these models requires knowledge of WIMP interactions with the heavy nuclei used in current detectors and light nuclei to be possibly used in next-generation searches~\cite{Hertel:2018aal}. Although the microscopic theory of WIMPs is unknown, the WIMP interactions with quarks and gluons can be parameterized at low-energy using EFT methods.  At operator dimensions six and seven, a number of WIMP-quark operators appear, involving in principle all quark bilinears (see Ref.~\cite{Bishara:2017pfq} and references therein).
To interpret the results of direct dark-matter searches and translate these into limits on dark-matter models, SM input is therefore necessary. By computing the appropriate single- and few-nucleon MEs, 
LQCD provides the needed non-perturbative bridge between the EFT description in terms of quarks and gluons  and the appropriate nucleon/nuclear EFTs and many-body methods~\cite{Fitzpatrick:2012ix,Cirigliano:2012pq,Menendez:2012tm,Hoferichter:2015ipa,Hoferichter2016,Korber:2017ery,Andreoli:2018etf}. LQCD can also be used to study the dynamics of new strongly-interacting theories that are proposed to govern the dark-matter sector. This latter class of investigations is discussed in the companion whitepaper on ``Lattice Gauge Theory for Physics Beyond the Standard Model''~\cite{Brower:2018qcd}, and will not be discussed here.

\
\

\noindent
\emph{Progress report:} While the up and down scalar MEs of the nucleon can be extracted from pion-nucleon scattering experiments, LQCD determinations are now competitive in terms of precision.\footnote{It is notable that recent LQCD determinations are in slight tension with the latest analyses of experimental data~\cite{RuizdeElvira:2017stg}.} Heavy-quark contributions to the MEs are amenable to perturbation theory, while LQCD is the key tool used to obtain the strange contributions. LQCD determinations of the tensor charge of the up and down quarks reach a precision of 3--7\%, whereas the corresponding strange charge is bounded to be substantially smaller~\cite{Gupta:2018lvp}.  The flavor-diagonal axial charge of the up and down quarks is now known to 5--8\% accuracy, and the smaller strange contribution is known at $\sim15$\%~\cite{Gupta:2018zkh}. Calculations also exist for the form-factors of the pseudoscalar density and the axial and vector currents, though all the systematics are not yet under control, see Ref.~\cite{Jang:2018djx, Bali:2018qus} and references therein. Calculations of both strange and light scalar MEs with the physical values of the quark masses have achieved 10\%-15\% precision in the last five years~\cite{Ren:2014vea,Yang:2015uis,Durr:2015dna,Bali:2016lvx,Abdel-Rehim:2016won}.

\begin{figure}[!t]
\centering
\includegraphics[width=1\columnwidth]{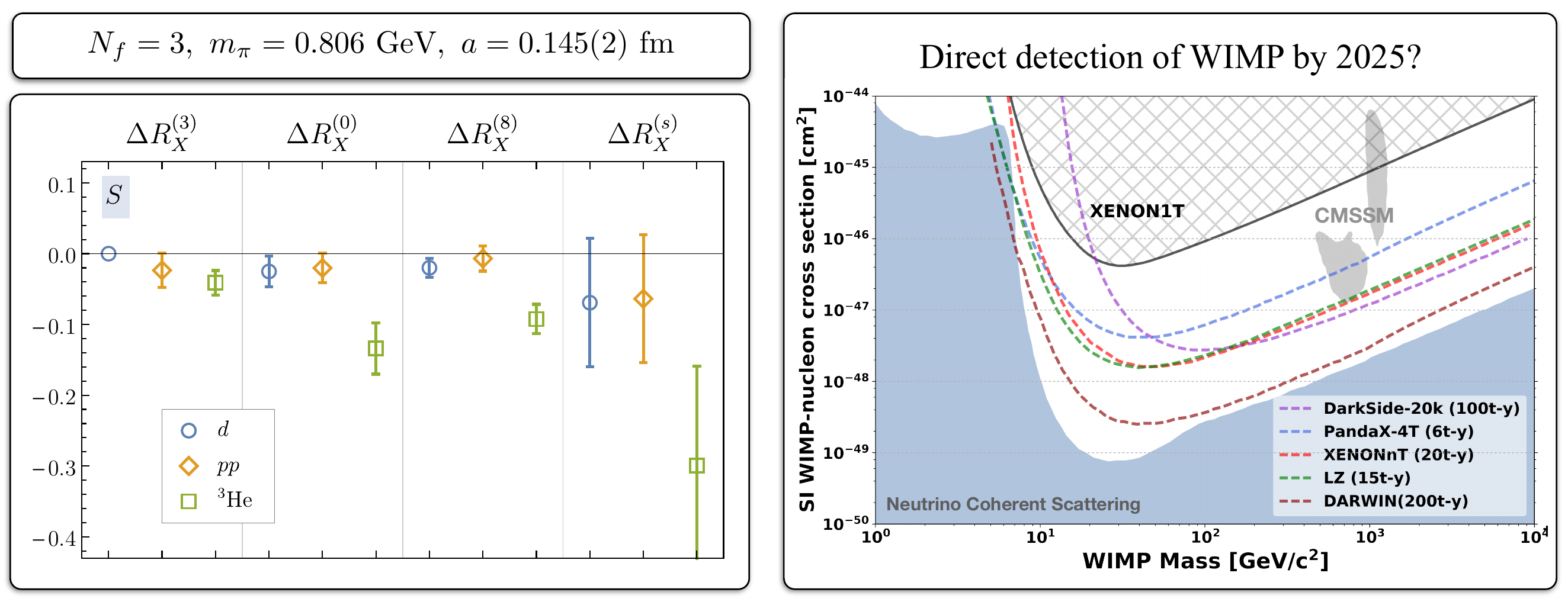}
\caption{The left panel (courtesy of the NPLQCD collaboration) shows deviation of the MEs of scalar, axial and tensor interactions in nuclei up to $A=3$ from their values in a noninteracting nucleon model, using LQCD albeit at an unphysically large quark masses. The quantities $\Delta R^{(3,0,8,s)}$ are defined in Ref.~\cite{Chang:2017eiq}. This investigation indicates a large ($\sim10\%$-level) nuclear effect for scalar interactions, which if it persists at physical values of the quark masses, can have a significant effect in MEs of scalar currents in large isotopes (a large quenching can be a possibility considering the quenching of the axial charge in such isotopes~\cite{PhysRevC.47.163}). Reliable theory input for scalar nuclear MEs will be crucial to interpret experimental searches in terms of the popular scalar portal scenarios. As is seen in the right panel (courtesy of Kaixuan Ni), the remaining not-yet-excluded region of the WIMP mass given the cross-section sensitivity of experiments will be considerably reduced in upcoming years, and reliable interpretation of the outcome of these searches will depend upon controlled nuclear MEs.}
\label{fig:scalarnplqcd}
\end{figure}

Ultimately, there is considerable uncertainty inherent in relating the scalar MEs of the nucleon to the scalar MEs of the nuclei used in direct searches for dark matter (e.g., Xenon with atomic number $A=131$), and controlled determinations of the nuclear MEs are of great interest. Recently, the first calculation of the light and strange quark sigma terms of light nuclei with $A\le 3$ was achieved~\cite{Chang:2017eiq}, albeit at larger-than-physical values of the quark masses. It is notable that the scalar charges of these light nuclei were found to be suppressed at the 10\% level relative to expectations for noninteracting nucleons, see the left panel of Fig.~\ref{fig:scalarnplqcd}. If this feature persists at the physical values of quark masses, it will be important that these effects not be neglected in the interpretation of direct-detection limits. As is seen in the right panel of Fig.~\ref{fig:scalarnplqcd}, the remaining not-yet-excluded region of the WIMP-nucleus cross section vs. the WIMP mass will be in reach of experiments planned for the next 5 years. To reliably convert limits on the cross sections from these experiments to a bound on the WIMP mass requires more precise knowledge of nuclear MEs.

\
\

\noindent
\emph{5-year goals and plans:}
\begin{itemize}
\item[--]{Few-percent level determinations of zero-recoil MEs of light and strange quark bilinear operators in the nucleon, including scalar and axial bilinears, can be categorized as \emph{straightforward}, given the emergence of Exascale computing resources in a few years.}
\item[--]{Given controlled calculations of the scalar MEs of light nuclei, these MEs in larger phe\-nom\-e\-nol\-o\-gic\-ally-rel\-e\-vant nuclei can be obtained by matching to chiral EFT calculations of few and many-body systems. A few-percent precision on MEs may be a requirement to isolate the effects beyond that expected from a naive impulse approximation, in which nuclear effects are ignored. While performing these calculations with physical quark masses is a computationally \emph{challenging} program, this goal may be achievable in the next 5 years.}
\item[--]{Beyond scalar portal models, there are of course many other possibilities for dark-matter interactions with the SM sector~\cite{Kumar:2013iva, Bishara:2017pfq}. For example, dark-matter interactions may be spin-dependent, and, depending on the dark-matter densities in our local environment, velocity-dependent couplings may also be important~\cite{Busoni:2017mhe}. Constraining these and other models requires knowledge of a range of QCD MEs, including those describing parton structure, which can be determined using LQCD. As for the scalar MEs, few-percent precision on the relevant nucleon MEs, and first controlled calculations of MEs in light nuclei, are possible on a short timescale, if particular needs are identified.}
\end{itemize}
%

   \section{Precision $\beta$ decay for searches of new physics}
  \label{sec:betadecay}
\noindent
\emph{Motivation:} Precision measurements of neutron and nuclear beta decay, at the 0.1\% level or better,  
provide a very competitive probe of new physics well into the LHC era~\cite{Gonzalez-Alonso:2018omy}.
This statement is best quantified using the EFT framework~\cite{Bhattacharya:2011qm,Cirigliano:2012ab,Gonzalez-Alonso:2018omy}, 
in which,  at dimension-six, five non-standard  effective couplings can be probed via beta decays 
(the five other couplings that involve the right-handed neutrinos appear in observables only quadratically, resulting in lower sensitivity).
Normalizing the effective Lagrangian to $G_F V_{ud}$,  these five dimensionless couplings parameterize  the various BSM couplings of the left-handed lepton current:
 a correction to the usual SM left-handed quark current operator ($\epsilon_L$),  a right-handed quark current operator ($\epsilon_R$),  
 and the three chirality-flipping  scalar, pseudoscalar, and  tensor quark operators ($\epsilon_{S,P,T}$). 

\
\

\noindent
\emph{Progress report and 5-year goals and plans:} LQCD provides key input in matching the quark-level effective Lagrangian to the hadronic and nuclear level, 
namely the so-called isovector nucleon charges $g_\Gamma$, defined by 
\begin{equation}
\langle p (p',s') |\bar{u} \Gamma d | n (p,s) \rangle = g_\Gamma \, \bar{u}_p (p's')  \Gamma u_n (p,s)~,
\end{equation}
for the five Dirac structures $\Gamma\equiv 1,\,\gamma_5,\,\gamma_\mu,\, \gamma_5\gamma_\mu\,$ and $[\gamma_\mu,\gamma_\nu]$.   
Knowledge of scalar and tensor charges at the  ${} < 10\%$ level makes the next generation precision beta decay 
experiments compete with the LHC~\cite{Bhattacharya:2011qm,Cirigliano:2012ab} in constraining  $\epsilon_{S,T}$ at the 
few ${} \times 10^{-4}$ level,  i.e., probing effective scales of new physics close to 10 TeV. 
While the desired accuracy on $g_{S,T}$  has been achieved by several 
groups~\cite{Gupta:2018qil,Alexandrou:2017qyt,Bali:2014nma,Green:2012ej,Aoki:2010xg}, 
a confirmation of this would be desirable and  would entail  relatively \emph{straightforward} calculations.   

Knowledge of the axial charge $g_A$ will improve the model-independent bound on possible right-handed currents~\cite{Bhattacharya:2011qm,Alioli:2017ces}, 
since beta-decay experiments are sensitive to the combination $\bar{g}_A = g_A  (1 - 2 \epsilon_R)$, where $\epsilon_R$ parameterizes non-standard right-handed quark currents. 
Lattice calculation of $g_A$ routinely achieve a precision at the few~\% 
level~\cite{Gupta:2018qil,Alexandrou:2017hac,Abdel-Rehim:2015owa,Bali:2014nma,Horsley:2013ayv,Capitani:2012gj, Green:2012ud, Bratt:2010jn, Lin:2008uz, Yamazaki:2008py} and a recent calculation has claimed a 1\% precision~\cite{Chang:2018uxx}.
Independent determinations at the percent-level, especially using different methods, are highly desirable, and are plausible with higher statistics.  
Future improvements resulting in determinations at the 0.2-0.3\% level, requiring \emph{extremely challenging} calculations, would start probing the  $W$-quark right-handed coupling $\epsilon_R$ 
at a level that competes with the best bounds on $\epsilon_L$  from the Z-pole measurements. 

Finally,  we mention the opportunity for LQCD 
 to contribute to improving the understanding of 
 radiative corrections in $\beta$ decays. 
This is of high interest in light of recent results of Ref.~\cite{Seng:2018yzq}, 
which differ significantly from the previously established results~\cite{Marciano:2005ec} 
(barring compensating changes to the nuclear effects, the new radiative correction result leads to a $\sim 4$-$\sigma$ deviation in the CKM unitarity test). 
LQCD can contribute in two ways: 
(i) First, it can provide indirect input by calculating of the time-ordered product of a weak and electromagnetic current 
(inserted with momenta $\pm q$)  between neutron and proton, at a number of fixed $q$'s. This would provide a calibration for some of the input going in the calculation of  Ref.~\cite{Seng:2018yzq}.  The evaluation of the four-point functions or alternative methods~\cite{Seng:2019plg} would be \emph{challenging}. 
(ii) Second, LQCD can provide a more direct input by computing the $O(\alpha)$ contribution to $n \to p e \bar{\nu}_e$, using techniques similar to the ones developed for the calculation of radiative corrections to  $K (\pi) \to  \ell \nu_\ell$~\cite{Giusti:2017dwk}. 
This calculation is likely to be \emph{extremely challenging}.

   \section{Isotope-shift spectroscopy}
  \label{sec:isotopshift}
\noindent
Limits on new physics leading to spin-independent interactions between the neutron and the electron can be constrained using optical frequency measurements and frequency shifts between isotopes of hydrogen and helium atoms, light ions including lithium and nitrogen~\cite{Delaunay:2017dku}, and heavy atoms and ions~\cite{Berengut:2017zuo,Delaunay:2016brc,Frugiuele:2016rii}.  
QCD constraints on the scalar MEs of the nucleon and of the relevant nuclei are important in these calculations; recent progress and the outlook for LQCD calculations of these quantities is detailed above in the discussion of WIMPs. Unlike for direct DM searches which use large nuclei, however, the scalar MEs for the $A\lesssim 4$ light nuclei which are relevant in~Ref.~\cite{Delaunay:2017dku} can be provided from LQCD with controlled uncertainties on a 5-year timescale. The charge radius differences between nuclear isotopes are also necessary input. Radii can be extracted precisely from scattering experiments, but can be contaminated by new physics at various levels; in the longer term, LQCD calculations of charge radii for $A\lesssim 4$ nuclei will be feasible and will provide valuable constraints.

  \section{Computing needs}
  \label{sec:computing}
\noindent
Calculations involving nucleons are particularly demanding due to the exponential growth of noise in nucleon correlation functions. The noise in the single-nucleon correlator at  large (Euclidean) times behaves as $\sim e^{-3 m_{\pi} t/2}$, while the signal goes as $\sim e^{ -m_N t}$. The signal-to-noise ratio, therefore, degrades as $\sim e^{ -(m_N -3m_{\pi}/2) t}$ given the small mass of the pions~\cite{Parisi:1983ae, Lepage:1989}. In recent years, methods have been proposed based on domain decompositions at the level of correlation functions~\cite{Wagman:2016bam, Wagman:2017xfh, Liu:2017man} and the quark propagators~\cite{Ce:2016idq, Ce:2016ajy}, aiming to alleviate the noise in correlation functions, or enable the signal to be extracted from noisy correlators. This remains an active area of research in the field, with the hope that deep understanding of, and novel technologies on how to form, LQCD correlation functions may eliminate the need for an exponential increase in computational resources to achieve a signal for observables, see the companion whitepaper on ``Status and Future Perspectives for Lattice Gauge Theory Calculations to the Exascale and Beyond''~\cite{Joo:2018qcd}.

For LQCD studies in the area of Fundamental Symmetries, many of the ultimate goals require MEs of operators between multi-hadron states. Pioneering calculations have been successfully performed for systems with a small number of nucleons ($A< 5$), but still with up/down quark masses heavier than those in nature~\cite{Beane:2012vq,Yamazaki:2012hi,Doi:2011gq,Orginos:2015aya,Yamazaki:2015asa,Berkowitz:2015eaa}. To gain statistical control and to avoid overwhelming systematic error from excited-state contamination, several approaches have been implemented, such as faster and more efficient quark propagator solvers, efficient codes optimized to target HPC architectures, statistical-error reduction techniques, and more efficient ways of performing quark contractions for multi-hadron systems. Together with the support of HPC resources and its rapid evolutions, these developments brought immense speedups in computations, enabling today's flourishing lattice computations (see a companion whitepaper for state-of-the-arts examples~\cite{Joo:2018qcd}). The USQCD collaboration will continue to make breakthroughs in the efficient use of valuable HPC resources to provide indispensable theoretical results.

The software requirements are common to all hadron computations in the subcategories in this whitepaper. The requirements on gauge-field ensembles needed for carrying out various types of calculations can be different. A few examples are:
\begin{itemize}
\item[--]{For EDM calculations, a devoted QCD ensemble generation with CP-violating QCD action, especially the one with the nonzero $\theta$ vacuum angle, will be necessary.}
\item[--]{For calculations of many new-physics related MEs, ensembles of gauge-field configurations with good chiral symmetry properties are highly valuable, as they substantially reduce the number of relevant effective operators with a given mass dimension.}
\item[--]{For calculations involving nuclei, with vastly varying intrinsic energy scales compared with the QCD scale (e.g., consider the simplest nucleus, a deuteron, an extremely shallow bound state of two nucleons with a binding of only $\sim 2.2$ MeV), ensembles with larger volumes are required. The dense finite-volume spectra will pose a challenge in identifying ground and excited states out of Euclidean correlation functions, and more sophisticated techniques based on a variational approach, such as those customary in the mesonic sector~\cite{Briceno:2017max} and recently in the baryonic sector~\cite{Francis:2018qch}, will be a necessity.}
\item[--]{For calculations of MEs involving multiple nucleons, ensembles with multiple volumes and ideally various boundary conditions are essential so that a Lellouch-L\"uscher-type formalism can be used to extract physical MEs~\cite{Lellouch:2000pv,Detmold:2004qn,Briceno:2012yi,Agadjanov:2014kha,Briceno:2014uqa,Briceno:2015csa,Briceno:2015tza}. One should also perform calculations with various momenta for the initial and final states and for the momenta inserted at the current so to have access to multiple kinematic points.}
\end{itemize}

Finally, it must be noted that LQCD in its current form, and with classical resources, will likely not be the optimal method to directly obtain the MEs of operators related to the Fundamental Symmetries program in heavy isotopes, and a combination of LQCD, EFT and nuclear many-body methods are required to impact experiments. Computational challenges associated with nuclear LQCD calculations are enumerated in other companion whitepapers~\cite{Detmold:2018qcd,Kronfeld:2018qcd,Joo:2018qcd}. In short, these challenges include:
\begin{itemize}
\item[]{
\emph{i) Complexity of nuclear correlation functions:} A quark-level nuclear correlation function requires \emph{Wick} contractions of $2N_p+N_n$ number of up quarks in the nuclear ``source'' in all possible ways with the same number of up quarks in the nuclear ``sink'', as well as contracting $N_p+2N_n$ number of down quarks with the same number of down quarks, for statistical averaging over vacuum gauge-field configurations to follow. Here, $N_p$ denotes the number of protons and $N_n$ is the number of neutrons in the multi-nucleon system under study. This means a factorial growth in the complexity of the computation as the atomic number increases. This problem can be ameliorated with clever techniques~\cite{Detmold:2012eu, Doi:2012xd}, but still presents a roadblock in first-principle QCD calculations of large isotopes relevant to the Fundamental Symmetries program.
}
\item[]{
\emph{iii) Signal degradation:} The exponential degradation of the signal discussed above for single nucleons occurs at much earlier times in the correlator if a system with atomic number $A$, with the signal-to-noise ratio scaling as $\sim e^{ -A(m_N -3m_{\pi}/2) t}$.
}
\item[]{
\emph{ii) Fine and dense density of states:} While the excitation gap of single-nucleon state is of the order of the QCD scale, those of a large nucleus are orders of magnitude smaller. This means that a LQCD calculation of a large nucleus should obtain its mass (could be hundreds of GeVs) with an incredible precision, so as to isolate the lowest states from their nearby excitations (where the energy gap could be only a few keVs). Given the severe signal-to-noise degradation and the substantial HPC resources required to overcome the problem, such a precision goal may not be within reach in the next decade without revolutionary paradigms in computing.
}
\end{itemize}
It is therefore important to emphasize that the LQCD effort in Fundamental Symmetries, which will be limited to single and few-nucleon observables in coming years, will not pay off without a cohesive and coordinated effort that involves EFT and nuclear many-body physicists that build the bridge between the QCD input for few-nucleon quantities and target quantities in the experimental program. Such effort has gained significant momentum in recent years~\cite{Barnea:2013uqa, Kirscher:2015yda, Contessi:2017rww, Kirscher:2017fqc, Bansal:2017pwn} given the success of the LQCD program in delivering results that were not possible in the past, and will continue to accelerate as LQCD studies of nucleons and nuclei mature~\cite{Detmold:2018qcd}.

\bibliography{bibi}

\end{document}